\documentclass[sigconf]{acmart}

\usepackage{algorithm}
\makeatletter
\renewcommand{\ALG@name}{Protocol}
\usepackage{algorithmic}

\usepackage{graphicx}

\usepackage{float}
\usepackage[algo2e]{algorithm2e}
\usepackage{amsmath, amsthm}
\usepackage{multirow}
\usepackage{dcolumn}
\newlength{\thinline}
\newlength{\thickline}
\usepackage{threeparttable}
\usepackage{esvect}
\usepackage{listings}
\usepackage{url}
\usepackage{subfigure}
\newtheorem{theorem}{Theorem}
\usepackage{booktabs}
\usepackage{tabularx}
\usepackage{balance}
\usepackage{enumitem}
\usepackage{hyperref}
\usepackage{cleveref}
\usepackage{tcolorbox}
\usepackage{amsmath}

\tcbuselibrary{skins, breakable, theorems}

\newtcbtheorem[auto counter]{protocol}{\textbf{Protocol}}%
  {enhanced,float,
    colback = white,
    attach boxed title to top left = {yshift = -3mm, xshift = 5mm},
    fonttitle = \sffamily\bfseries}{pro}

\newtcbtheorem[auto counter]{funct}{\textbf{Functionality}}%
  {enhanced,float,
    colback = white,
    attach boxed title to top left = {yshift = -3mm, xshift = 5mm},
    fonttitle = \sffamily\bfseries}{funct}
    
\newtcbtheorem[auto counter]{simulator}{\textbf{Simulator}}%
  {enhanced,float,
    colback = white,
    attach boxed title to top left = {yshift = -3mm, xshift = 5mm},
    fonttitle = \sffamily\bfseries}{simulator}

\newcommand{\share}[1]{\langle {#1} \rangle}
\newcommand{\additiveshare}[1]{\llbracket{#1}\rrbracket }

\newcommand{\fname}{\texttt{Ents}\xspace}
\newcommand{\ekk}{\mathscr{k}}

\usepackage{dblfloatfix}

\AtBeginDocument{%
  \providecommand\BibTeX{{%
    \normalfont B\kern-0.5em{\scshape i\kern-0.25em b}\kern-0.8em\TeX}}}

\copyrightyear{2024}
\acmYear{2024}
\setcopyright{acmlicensed}
\acmConference[CCS '24] {Proceedings of the 2024 ACM SIGSAC Conference on Computer and Communications Security}{October 14--18, 2024}{Salt Lake City, UT, USA.}
\acmBooktitle{Proceedings of the 2024 ACM SIGSAC Conference on Computer and Communications Security (CCS '24), October 14--18, 2024, Salt Lake City, UT, USA}
\acmISBN{979-8-4007-0636-3/24/10}
\acmDOI{10.1145/3658644.3670274}

\begin{document}

\author{
 Guopeng Lin$^\$$, Weili Han$^\$$, Wenqiang Ruan$^\$$, Ruisheng Zhou$^\$$, Lushan Song$^\$$, Bingshuai Li$^+$, Yunfeng Shao$^+$
}

\affiliation{%
  \institution{$^\$$Laboratory for Data Security and Governance, Fudan University, $^+$Huawei Technologies}
  \country{China}
}




\title{ \fname: An Efficient Three-party Training Framework for Decision Trees by Communication Optimization}

\begin{abstract}

Multi-party training frameworks for decision trees based on secure multi-party computation enable multiple parties to train high-performance models on distributed private data with privacy preservation. 
The training process essentially involves frequent dataset splitting according to the splitting criterion (e.g. Gini impurity). However, existing multi-party training frameworks for decision trees demonstrate communication inefficiency due to the following issues:  (1) They suffer from huge communication overhead in securely splitting a dataset with continuous attributes. (2) They suffer from huge communication overhead due to performing almost all the computations on a large ring to accommodate the secure computations for the splitting criterion.

In this paper, we are motivated to present an efficient three-party training framework, namely \fname, for decision trees by communication optimization.   For the first issue,  we present a series of training protocols based on the secure radix sort protocols~\cite{hamada2021efficient} to efficiently and securely split a dataset with continuous attributes. For the second issue,  we propose an efficient share conversion protocol to convert shares between a small ring and a large ring to reduce the communication overhead incurred by performing almost all the computations on a large ring. Experimental results from eight widely used datasets show that \fname outperforms state-of-the-art frameworks by $5.5\times \sim 9.3\times$ in communication sizes and $3.9\times \sim 5.3\times$ in communication rounds. In terms of training time,
\fname yields an improvement of $3.5\times \sim 6.7\times$.  To demonstrate its practicality, \fname requires less than three hours to securely train a decision tree on a widely used real-world dataset (Skin Segmentation) with more than 245,000 samples in the WAN setting.

\end{abstract}




\settopmatter{printfolios=true}

\maketitle

\section{Introduction} \label{sec.intro}

Decision trees are one of the most popular machine learning models~\cite{kaggle2021survey, kaggle2020survey}. They are widely used in practical applications such as medical diagnosis~\cite{ huyut2022prediction, tyagi2022prediction, sciaviccovoice} and stock forecasting~\cite{Chen2021Forecasting, lee2022predictive, arya2022stock} due to their interpretability, which is very important in medical and financial scenarios. In these domains, black-box machine-learning models, such as neural networks, might be limited. For example, in medical diagnosis, doctors usually prefer a predictive model that not only performs well but also can identify the key biological factors directly affecting the patient's health outcome.

The accuracy of decision trees largely depends on the number of high-quality training data, which are usually distributed among different parties.  For instance, in the medical diagnosis for a rare disease, each hospital usually only has several cases. Thus, the data of a hospital is usually not enough to train an acceptable model. As a result,  multiple hospitals would like to collaboratively train an accurate diagnosis decision tree. 
However, in this case, the multiple hospitals could not directly share their private data of the rare disease to train a decision tree due to released privacy protection regulations and laws (e.g. GDPR~\cite{voigt2017eu}). 

Multi-party training frameworks~\cite{hamada2021efficient, MarkAbspoel2021SecureTO, deHoogh2014Practical, liu2020towards} for decision trees based on secure multi-party computation (MPC for short) are proposed to resolve the above issue. However, these frameworks suffer from practical scenarios due to two communication inefficiency issues as follows: (1) Existing multi-party training frameworks for decision trees suffer from huge communication overhead in securely splitting a dataset with continuous
attributes. Continuous attributes, e.g. salary, commonly exist in real-world datasets. However, securely splitting such datasets requires a vast number of secure comparison operations, resulting in huge communication overhead. Abspoel et al.~\cite{MarkAbspoel2021SecureTO} propose securely generating permutations from continuous attributes at the beginning of the training process. These pre-generated permutations could be used to securely sort the training samples according to corresponding attributes, so that a lot of secure comparison operations could be saved. Nevertheless, their method requires securely training each decision tree node with a padded dataset to hide the splitting information, leading to huge communication overhead that grows exponentially as the height of the decision tree increases. To limit the communication overhead to grow linearly as the tree height increases, Hamada et al.~\cite{hamada2021efficient} propose group-wise protocols to securely train each decision tree node with a non-padded dataset while keeping the splitting information private. However, the group-wise protocols are incompatible with the pre-generated permutations. As a result, Hamada et al.'s method requires the repeated secure generation of permutations, again resulting in huge communication overhead.  (2) Existing multi-party training frameworks for decision trees suffer from huge communication overhead due to performing almost all computations on a large ring to accommodate the secure computations for the splitting criterion (e.g. Gini impurity). The secure computations for the splitting criterion are a crucial step in the secure training process of decision trees and usually involve several sequential bit-length expansion operations, such as secure multiplication and division operations. These operations often produce intermediate or final results that require twice as many bits for representation compared to their inputs. Performing these operations sequentially several times requires a large ring to represent the intermediate or final results. Therefore, most existing multi-party training frameworks for decision trees, such as~\cite{deHoogh2014Practical, MarkAbspoel2021SecureTO, hamada2021efficient}, conduct almost all computations on a large ring, leading to huge communication overhead.


In this paper, we are motivated to propose an efficient three-party training framework, namely \fname.  \fname enables three parties, such as the hospitals mentioned above, to efficiently train a decision tree while preserving their data privacy. In \fname, we propose two optimizations to reduce the communication overhead led by the above issues.  (1) We propose a series of training protocols based on the secure radix sort protocols~\cite{hamada2021efficient} to efficiently and securely split a dataset with continuous attributes. One of the training protocols securely updates the pre-generated permutations to be compatible with the group-wise protocols, and the other protocols leverage the updated permutations and the group-wise protocols to securely split the dataset (train the layers of a decision tree). As a result, \fname can securely train a decision tree with only linearly growing communication overhead, while eliminating the need to repeatedly generate permutations. (2) We propose an efficient share conversion protocol to convert shares between a small ring and a large ring to reduce the communication overhead incurred by performing almost all computations on a large ring. We observe that though the secure computations for the splitting criterion usually involve bit-length expansion operations like secure multiplication and division operations, other computations in the training process just involve bit-length invariant operations, such as secure addition and comparison operations, which at most cause a single-bit increase. While the bit-length expansion operations should be performed on a large ring, the bit-length invariant operations can be performed on a small ring. Therefore,  we can employ our proposed efficient share conversion protocol to convert shares to a large ring when performing the bit-length expansion operations and then convert the shares back to a small ring to reduce the communication overhead of the bit-length invariant operations.

We summarize the main contributions in \fname as follows:

\begin{itemize}[leftmargin=*]

\item  Based
on the secure radix sort protocols, we propose a series of efficient training protocols for decision trees. 

\item We propose an efficient share conversion protocol to convert shares between a small ring and a large ring. 

\end{itemize}

To evaluate the efficiency of \fname~\footnote{We opened our implementation codes at https://github.com/FudanMPL/Garnet/tree/Ents.}, we compare \fname against two state-of-the-art three-party training frameworks~\cite{hamada2021efficient, MarkAbspoel2021SecureTO} for decision trees with eight widely-used real-world datasets from the UCI repository~\cite{kelly2023uci}. The experimental results show that \fname outperforms these two frameworks by $5.5\times \sim 9.3\times$ in communication sizes, $3.9\times \sim 5.3\times$ in communication rounds, and $3.5\times \sim 6.7\times$ in training time. Notably, \fname requires less than three hours to train a decision tree on a real-world dataset (Skin Segmentation) with more than 245,000 samples
in the WAN setting. These results show that \fname is promising in the practical usage (i.e. industrial deployment) of privacy preserving training for decision trees.

\section{Preliminaries}
\label{sec.pre}

\subsection{Decision Tree Training}


\noindent \textbf{Structure of decision trees.} Given a training dataset $\mathcal{D}$ containing $n$ samples, each of which consists of $m$ continuous attributes $a_0$,$\cdots$,$a_{m-1}$ and a label $y$, a decision tree built on this dataset  $\mathcal{D}$ is usually a binary tree. Each internal node of the decision tree includes a split attribute index $i$ ($i \in [0, \cdots,m-1])$ and a split threshold $t$. The split attribute index $i$ and the split threshold $t$ composite a split point $(a_i, t): a_i < t$, which is used for splitting datasets during the training process.  Each leaf node of the decision tree includes a predicted label $y_{\mathit{pred}}$.

\noindent \textbf{Training process for decision trees.} A decision tree is usually trained from the root node to the leaf nodes using a top-down method. For training each node, the first step is to check whether the split-stopping condition is met. In this paper, we adopt the commonly used split-stopping condition~\cite{akavia2022privacy, MarkAbspoel2021SecureTO, hamada2021efficient}, i.e. when the current node's height reaches a predefined tree's height $h$. If the split-stopping condition is met, the current node becomes a leaf node. Its predicted label $y_{pred}$ is set to the most common label in its training dataset $\mathcal{D}^{node}$, which is a subset of  $\mathcal{D}$.  Otherwise, the current node becomes an internal node, and its split point is computed according to the chosen splitting criterion, such as the modified Gini impurity mentioned in Section~\ref{sec.criterion}. Assuming the split point is $(a_i, t)$,  the training dataset $\mathcal{D}^{node}$ of the current node is split into two sub-datasets, $\mathcal{D}^{node}_{a_i < t}$ and $\mathcal{D}^{node}_{a_i \geq t}$. Here, $\mathcal{D}^{node}_{a_i < t}$  refers to a dataset that contains all samples whose $i$-th attribute values are smaller than $t$ in $\mathcal{D}^{node}$, and $\mathcal{D}^{node}_{a_i \geq t}$  refers to a dataset that contains all samples whose $i$-th attribute values are greater than or equal to $t$ in $\mathcal{D}^{node}$. Then, the left child node and right child node of the current node are trained using $\mathcal{D}^{node}_{a_i < t}$ and $\mathcal{D}^{node}_{a_i \geq t}$, respectively.

\subsection{Splitting Criterion}
\label{sec.criterion}
Several splitting criteria have been proposed, such as information gain~\cite{quinlan1986induction}, information gain ratio~\cite{quinlan2014c4}, and Gini impurity~\cite{loh2011classification}. In this paper, we follow the previous studies~\cite{hamada2021efficient, MarkAbspoel2021SecureTO} to use the modified Gini impurity, which is easy to compute in MPC~\cite{hamada2021efficient, MarkAbspoel2021SecureTO}, as the splitting criterion to implement our proposed \fname.


\noindent\textbf{Modified Gini impurity.} 
Let $v$ be the number of labels, and a label $y \in [0, \cdots, v-1]$,  the modified Gini impurity for a split point $(a_i, t)$ in a dataset $\mathcal{D}$ is defined as follows:

\begin{equation}
\label{eq.modifiedgini}
    \begin{aligned}
        \mathit{Gini}_{a_i < t}(\mathcal{D}) &= \frac{\sum_{l=0}^{v-1}|\mathcal{D}_{a_i < t  \wedge y=l}|^2}{|\mathcal{D}_{a_i < t}|} \\
&+ \frac{\sum_{l=0}^{v-1}|\mathcal{D}_{a_i \geq t  \wedge y=l}|^2}{|\mathcal{D}_{a_i \geq t}|}
    \end{aligned}
\end{equation}
  $\mathcal{D}_{a_i < t  \wedge y=l}$ refers to a dataset that contains all samples whose $i$-th attribute values are smaller than $t$ and labels are $l$ in $\mathcal{D}$.  $\mathcal{D}_{a_i \geq t  \wedge y=l}$ refers to a dataset that contains all samples whose $i$-th attribute values are greater than or equal to $t$ and labels are $l$ in $\mathcal{D}$. Besides, $|\mathcal{D}_{a_i < t}|$, $|\mathcal{D}_{a_i \geq t}|$, $|\mathcal{D}_{a_i < t  \wedge y=l}|$ and $|\mathcal{D}_{a_i \geq t  \wedge y=l}|$ refer to the number of the samples in $\mathcal{D}_{a_i < t}$, $\mathcal{D}_{a_i \geq t}$, $\mathcal{D}_{a_i < t  \wedge y=l}$  and $\mathcal{D}_{a_i \geq t  \wedge y=l}$, respectively. 

During the training process, the split point $(a_i, t)$ of an internal node should be set to the one with the maximum modified Gini impurity $\mathit{Gini}_{a_i < t}(\mathcal{D}^{node})$~\cite{hamada2021efficient, MarkAbspoel2021SecureTO}, where $\mathcal{D}^{node}$ refers to the training dataset of the node.

\subsection{Secure Multi-party Computation}
\label{sec.mpc}

MPC enables multiple parties to cooperatively compute a function while keeping their input data private. There are several technical routes of MPC,  including secret sharing-based protocols~\cite{mohassel2018aby3}, homomorphic encryption-based protocols~\cite{dulek2016quantum}, and garbled circuit-based protocols~\cite{ciampi2021threshold}. In this paper, we adopt the three-party replicated secret sharing (RSS for short) as the underlying protocol of \fname for its efficiency and promising security.

In this section, we first introduce the three-party replicated secret sharing. Then, we sequentially introduce the basic operations~\cite{mohassel2018aby3, tan2021cryptgpu}, the vector max protocol~\cite{hamada2021efficient}, the secure radix sort protocols~\cite{chida2019efficient}, and the group-wise protocols~\cite{hamada2021efficient} based on RSS, which are used in our proposed training protocols in \fname.

\subsubsection{Three-party Replicated Secret Sharing}
To share a secret $x$ on a ring $\mathbf{Z}_{2^\ekk}$ of size $2^\ekk$ using RSS to three parties $P_0$, $P_1$ and $P_2$, a party first generates three random numbers $\additiveshare{x}^0$, $\additiveshare{x}^1$ and $\additiveshare{x}^2$, where $x = (\additiveshare{x}^0 + \additiveshare{x}^1 + \additiveshare{x}^2) \ mod \ 2^\ekk$.
Then the party sends  $\additiveshare{x}^i$, $\additiveshare{x}^{i+1}$ (where $\additiveshare{x}^{2+1}$ refers to $ \additiveshare{x}^{0}$) to $P_i$.  In this paper, we use the notation $\share{x}$ to denote the replicated shares of $x$, which means $P_i$ holds $\share{x}^i$ $=$ $(\additiveshare{x}^i, \additiveshare{x}^{i+1})$.

To share a vector $\vec{x}$  using RSS, the three parties leverage the above way to share each element of $\vec{x}$. For example, given a vector $\vec{x} = [2,3,1]$, the secret-shared vector $\share{\vec{x}}$ would be $[\share{2}$,$ \share{3}$,$ \share{1}]$.

\subsubsection{Basic Operations Based on RSS}

 Let $c$ be a public constant, $\share{x}$ and $\share{y}$ be the replicated shares of $x$ and $y$, the basic operations~\cite{mohassel2018aby3, tan2021cryptgpu} based on RSS used in this paper is as follows:

\begin{itemize}

  \item \textbf{Secure addition with constant.} $\share{z} = \share{x} + c$, such that $z = x + c$.
  \item \textbf{Secure multiplication with constant.}  $\share{z} = c * \share{x} $, such that $z = c * x$.
  \item \textbf{Secure addition.} $\share{z} =   \share{x} + \share{y}$ , such that $z = x + y$.

    \item \textbf{Secure multiplication.} $\share{z} =  \share{x} * \share{y}$ , such that $z = x * y$.

    \item \textbf{Secure probabilistic truncation.} $\share{z} =   \mathit{Trunc}(\share{x}, c)$ , such that $z = \lfloor {x / 2^c}  \rfloor + \mathit{bit}$, $\mathit{bit} \in [-1, 0, 1]$.
    
    \item \textbf{Secure comparison.} $\share{z} =  \share{x} < \share{y}$ , such that if $x < y$, $z = 1$, otherwise $z = 0$.
    
    \item \textbf{Secure equality test.} $\share{z} =  \share{x} \overset{?}{=} \share{y}$ , such that if $x=y$, $z = 1$, otherwise $z = 0$.

    \item \textbf{Secure division.} $\share{z} =   \share{x} \ / \ \share{y}$ , such that $z = x / y$.
\end{itemize}

\subsubsection{Vector Max Protocol Based on RSS}

\label{sec.vectmax}
The protocol \textit{VectMax}~\cite{hamada2021efficient} is used to securely find an element in a vector whose position matches that of a maximum value in another vector. In this protocol, the parties input two secret-shared vectors $\share{\vec{x}}$ and $\share{\vec{y}}$, both of which have the same size. The parties get a secret-shared element $\share{z}$ as the output, such that $z = \vec{y}[i]$ when $\vec{x}[i]$ is a maximum value in $\vec{x}$ (if there are multiple maximum values, the last one is chosen). For example, given $\vec{x} = [2, 3, 1]$ and $\vec{y} = [4, 5, 6]$, $\mathit{VectMax}(\share{\vec{x}},$ $ \share{\vec{y}})$ $= \share{5}$. 


\subsubsection{Secure Radix Sort Protocols Based on RSS}
\label{sec.sort}
The secure radix sort protocols are based on a data structure called a permutation. A permutation $\vec{\pi}$ is a special vector whose elements are distinct from each other and belong to $[0,  n-1]$, where $n$ is the size of the permutation. A permutation $\vec{\pi}$ could be used to reorder elements of a vector whose size is the same as the permutation.  For example, given a permutation $\vec{\pi} = [3, 2, 4, 0, 1]$ and a vector $\vec{x} = [4, 9, 2, 9, 3]$, we can reorder the elements of $\vec{x}$ according to $\vec{\pi}$ to get a new vector $\vec{z}$ ($\vec{z}[\vec{\pi}[i]] = \vec{x}[i]$). Consequently, we obtain $\vec{z} = [9, 3, 9, 4, 2]$.

In this paper, we adopt the secure radix sort protocols proposed by Chida et al.~\cite{chida2019efficient}. To securely radix sort a secret-shared vector $\share{\vec{x}}$, the parties use two protocols: \textit{GenPerm} and \textit{ApplyPerm}. In the protocol \textit{GenPerm}, the parties input a secret-shared vector $\share{\vec{x}}$, and get a secret-shared permutation $\share{\vec{\pi}}$ as the output,  where $\vec{\pi}[i]$ represents the index of $\vec{x}[i]$ in ascending order. Note that this ascending order is stable. For example, given $\share{\vec{x}} = [\share{4}, \share{9}, \share{2}, \share{9}, \share{3}]$,  $\share{\vec{\pi}} =$ $ \mathit{GenPerm}(\share{\vec{x}}) = $ $[\share{2}, \share{3}, \share{0}, \share{4}, \share{1}]$.   In the protocol \textit{ApplyPerm}, the parties input a secret-shared permutation $\share{\vec{\pi}}$ and a secret-shared vector $\share{\vec{x}}$, and they get a secret-shared vector $\share{\vec{z}}$ that is created by securely reordering $\share{\vec{x}}$ according to $\share{\vec{\pi}}$ as the output. For example, given $\share{\vec{\pi}} = [\share{2}, \share{3}, \share{0}, \share{4}, \share{1}]$ and $\share{\vec{x}} = [\share{4}, \share{9}, \share{2}, \share{9}, \share{3}]$,  $\share{\vec{z}} =$ $ \mathit{ApplyPerm}(\share{\vec{\pi}}, \share{\vec{x}}) =$ $ [\share{2}, \share{3}, \share{4}, \share{9}, \share{9}]$. Thus, by first generating a secret-shared permutation $\share{\vec{\pi}}$  from a secret-shared vector $\share{\vec{x}}$ using the protocol \textit{GenPerm} and then applying $\share{\vec{\pi}}$ to $\share{\vec{x}}$ using the protocol \textit{ApplyPerm}, the parties can securely radix sort $\share{\vec{x}}$. Additionally, if the parties apply $\share{\vec{\pi}}$ to another secret-shared vector $\share{\vec{y}}$ with the same size of $\share{\vec{x}}$, they can securely radix sort $\share{\vec{y}}$ according to $\share{\vec{x}}$.


We detail the implementation of the protocol \textit{GenPerm}, because this protocol inspires our proposed permutation updating protocol \textit{UpdatePerms} (Protocol~\ref{pro:updateperms} in Section~\ref{sec.update_permuation}). \textit{GenPerm} is implemented with four subprotocols \textit{BitVecDec}, \textit{ApplyPerm}, \textit{GenPermByBit}, and \textit{ComposePerms}. In the protocol \textit{BitVecDec}, the parties input a secret-shared vector $\share{\vec{x}}$, and they get $\ekk$ secret-shared bit vectors $\share{\vec{b_0}}, \cdots, \share{\vec{b}_{\ekk-1}}$ as the output,  such that  $\vec{x}[i] = \sum_{j=0}^{\ekk-1}2^j*\vec{b_j}[i]$. Here, $\ekk$ is the bit length of the ring $\mathbb{Z}_{2^\ekk}$ on which the protocol \textit{BitVecDec} is performed. In the protocol \textit{GenPermByBit}, the parties input a secret-shared bit vector $\share{\vec{b}}$, and they get a secret-shared permutation $\share{\vec{\pi}}$ as the output, such that  $\vec{\pi}[i]$ represents the index of $\vec{b}[i]$ in ascending order. Note that this ascending order is also stable. In the protocol \textit{ComposePerms}, the parties input two secret-shared permutations $\share{\vec{\alpha}}$ and $\share{\vec{\beta}}$, and get a secret-shared permutation $\share{\vec{\pi}}$ as the output, such that $\share{\vec{\pi}}$ compose the effects of both $\share{\vec{\alpha}}$ and $\share{\vec{\beta}}$, i.e. $ \mathit{ApplyPerm}(\share{\vec{\pi}}, \share{\vec{x}})$ $ = $ $\mathit{ApplyPerm}$ $ 
(\share{\vec{\beta}}$, $\mathit{ApplyPerm}(\share{\vec{\alpha}}$, $\share{\vec{x}}))$.

As is shown in Protocol~\ref{pro:genperm}, the protocol \textit{GenPerm} consists of three stages. The first stage (Line 1): the parties use the protocol \textit{BitVecDec} to decompose the input $\share{\vec{x}}$ into $\ekk$ secret-shared bit vectors, $\share{\vec{b_0}}, \cdots, \share{\vec{b}_{\ekk-1}}$. The second stage (Line 2): the parties generate a secret-shared permutation $\share{\vec{\pi}}$ from the least significant bit vector $\share{\vec{b_0}}$. The third stage (Line 3-7): the parties update $\share{\vec{\pi}}$ with each remaining secret-shared bit vector. To  update $\share{\vec{\pi}}$ with the $i$-th secret-shared bit vector $\share{\vec{b_i}}$, the parties first apply $\share{\vec{\pi}}$ to $\share{\vec{b_i}}$ to make the $i$-th bits sorted according to the least significant $i-1$ bits, producing a new secret-shared bit vector $\share{\vec{b'}_i}$ (Line 4). Next, the parties generate a new secret-shared permutation $\share{\vec{\alpha}}$ from $\share{\vec{b'}_i}$ (Line 5), such that $\share{\vec{\alpha}}$ can be used to sort a secret-shared vector that has been sorted according to the least significant $i-1$ bits further according to the $i$-th bits. Finally, the parties use the protocol \textit{ComposePerms} compose the effect of $\share{\vec{\pi}}$ and $\share{\vec{\alpha}}$ to obtain the updated $\share{\vec{\pi}}$, which can be used to directly sort a secret-shared vector according to the least significant $i$ bits (Line 6).

Besides, we introduce another protocol \textit{UnApplyPerm}, which is also used in our proposed training protocols. In this protocol, the parties input a secret-shared permutation $\share{\vec{\pi}}$  and a secret-shared vector $\share{\vec{x}}$, and they get a secret-shared vector $\share{\vec{z}}$, which is generated by securely reversing the effect of applying $\share{\vec{\pi}}$ to $\share{\vec{x}}$, as the output. For example, given $\share{\vec{\pi}} = [\share{2}, \share{3}, \share{0}, \share{4}, \share{1}]$ and $\share{\vec{x}} = [\share{2}, \share{3}, \share{4}, \share{9}, \share{9}]$,  $\share{\vec{z}} = $ $\mathit{UnApplyPerm}(\share{\vec{\pi}}, \share{\vec{x}}) = $ $[\share{4}, \share{9}, \share{2}, \share{9}, \share{3}]$.


\begin{algorithm}
    \LinesNumbered
    \small
    \caption{$GenPerm(\share{\vec{x}})$}
    \label{pro:genperm}
    \begin{flushleft}
    \textbf{Input:}   A secret-shared vector $\share{\vec{x}}$.\\
    \textbf{Output:}  A secret-shared permutation $\share{\vec{\pi}}$.
    \end{flushleft}
    \begin{algorithmic}[1]
    \STATE $\share{\vec{b_0}}, \cdots, \share{\vec{b}_{\ekk-1}} = \mathit{BitVecDec}(\share{\vec{x}})$. 
    \STATE $\share{\vec{\pi}} = \mathit{GenPermByBit}(\share{\vec{b_0}})$.
    \FOR{$ i = 1\ to\ \ekk-1$ }  
        \STATE $\share{\vec{b'}_i} = \mathit{ApplyPerm}(\share{\vec{\pi}}$, $\share{\vec{b_i}})$. 
        \STATE $\share{\vec{\alpha}} = \mathit{GenPermByBit}(\share{\vec{b'}_i})$. 
        \STATE $\share{\vec{\pi}} = \mathit{ComposePerms}(\share{\vec{\pi}},\share{\vec{\alpha}})$. 
    \ENDFOR
 \end{algorithmic}
\end{algorithm}

\subsubsection{Group-Wise Protocols Based on RSS}
\label{sec.group}
We present the design of group-wise protocols proposed by Hamada et al.~\cite{hamada2021efficient}.   The fundamental data structure of the group-wise protocols consists of a secret-shared element vector $\share{\vec{x}}$ and a secret-shared group flag vector $\share{\vec{g}}$. The secret-shared group flag vector   $\share{\vec{g}}$ privately indicates the group boundaries of $\share{\vec{x}}$. Concretely, $\share{\vec{g}[i]} = \share{1}$ if the $i$-th element of $\share{\vec{x}}$ is the first element of a group, and $ \share{\vec{g}[i]} = \share{0}$ otherwise. According to this definition, $\share{\vec{g}[0]}$ always equals to $\share{1}$. For example, $\share{\vec{g}} = [\share{1}, \share{0}, \share{1}, \share{1}, \share{0}, \share{0}]$ and $\share{\vec{x}} = [\share{4}, \share{3}, \share{2}, \share{8}, \share{9}, \share{0}]$ mean that the six elements of $\share{\vec{x}}$ are divided into three groups:  $\{\share{4},\share{3}\}$, $\{\share{2}\}$, $\{\share{8},\share{9},\share{0}\}$. With this data structure, the elements of a secret-shared vector $\share{\vec{x}}$ can be divided into several groups without revealing any information about group membership.

\begin{figure}[htp]
    \centering
    \includegraphics[width=0.4\textwidth]{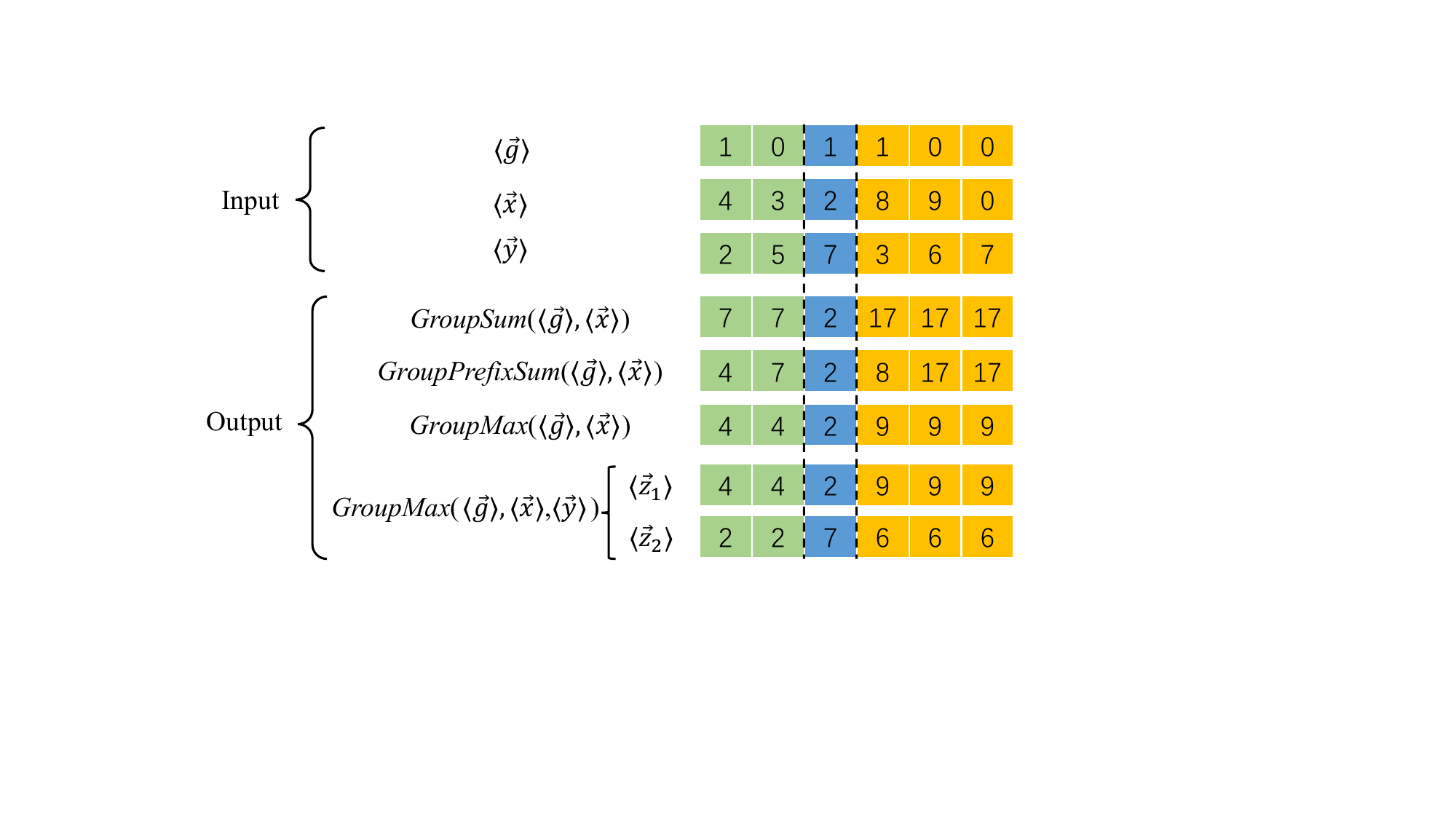}

    \caption{Examples of group-wise protocols. We use both dashed lines and colors to distinctly delineate the group boundaries.  }
  
    \label{group_examples}
\end{figure}

The group-wise protocols based on the above data structure include \textit{GroupSum}, \textit{GroupPrefixSum}, and \textit{GroupMax}. In all these protocols, the parties input a secret-shared group flag vector $\share{\vec{g}}$  and a secret-shared element vector $\share{\vec{x}}$, and get a secret-shared vector $\share{\vec{z}}$ as the output. The secret-shared vector $\share{\vec{g}}$, $\share{\vec{x}}$ and $\share{\vec{z}}$ are of the same size.  Here, we define $f(i)$ as the first element index of the group that $\vec{x}[i]$ belongs to, and $l(i)$ as the last element index of the group that $\vec{x}[i]$ belongs to. As is shown in Figure~\ref{group_examples}, the protocol \textit{GroupSum} securely assigns $\vec{z}[i]$ as the sum of the group that $\vec{x}[i]$ belongs to, i.e. $\vec{z}[i] = \sum_{j = f(i)}^{l(i)}\vec{x}[j] $. The protocol \textit{GroupPrefixSum} securely assigns $\vec{z}[i]$ as the prefix sum of the group that $\vec{x}[i]$ belongs to, i.e. $\vec{z}[i] = \sum_{j = f(i)}^{i}\vec{x}[j]$. The protocol \textit{GroupMax} securely assigns $\vec{z}[i]$ as the maximum of the group that $\vec{x}[i]$ belongs to, i.e. $\vec{z}[i] = \max_{ j \in [f(i),l(i)]}\vec{x}[j]$.  Besides, we introduce a extension version of the protocol \textit{GroupMax}, which are denoted as \textit{GroupMax}($\share{\vec{g}}$, $\share{\vec{x}}$, $\share{\vec{y}}$). In this protocol, the parties input a secret-shared group flag vector $\share{\vec{g}}$, a secret-shared vector $\share{\vec{x}}$, and a secret-shared vector $\share{\vec{y}}$, and they get two secret-shared vectors $\share{\vec{z}_1}$, $\share{\vec{z}_2}$ as the output, where $\vec{z}_1[i] = \vec{x}[k]$  and $\vec{z}_2[i] = \vec{y}[k]$ when $k = \arg\max_{ j \in [f(i),l(i)]}\vec{x}[j]$. The secret-shared vectors $\share{\vec{g}}$, $\share{\vec{x}}$, $\share{\vec{y}}$, $\share{\vec{z}_1}$ and $\share{\vec{z}_2}$ are also of the same size. 


\section{Overview of \fname}
\label{sec.overview}

\subsection{Architecture and Security Model}

\label{sec.arch}
The architecture of \fname involves three parties, each of which could be a corporation, organization, or even individual. They seek to collaboratively train a decision tree on their own data while preserving the privacy of their data.   The secure training process of \fname is as follows:   (1) the parties securely share their data with each other using RSS, ensuring that each party holds the shares of the entire training dataset. (2) the parties perform secure training protocols on this secret-shared training dataset to obtain the shares of a trained decision tree.  (3) the parties have the option to either send the shares of the trained decision tree to a designated party for tree reconstruction or utilize the shares for secure prediction on other samples.

\fname employs a three-party semi-honest security model with an honest majority, which is much more practical and widely used in the field of privacy preserving machine learning~\cite{tan2021cryptgpu, ruan2023private, hamada2021efficient, MarkAbspoel2021SecureTO, deHoogh2014Practical}, due to its efficiency and promising security. In this security model, the parties mentioned above will correctly execute the training protocols without colluding with each other. Besides, we also present the technical routines in Section~\ref{sec.diss}  for the two-party semi-honest security model with a dishonest majority. 


\subsection{Data Representation}

\label{sec.data_represent}

In our proposed training protocols, we add a subscript $M$ to $\share{x}$, producing $\share{x}_{M}$, to indicate the size of the ring on which a secret $x$ is shared, where $M = 2$ or $2 ^\ekk$ or $2^\ell$ ($2 < \ekk < \ell)$.

For a training dataset consisting of $n$ samples, each of which consists of $m$ continuous attributes and a label, we represent the dataset using $m$ secret-shared attribute vectors $\share{\Vec{a}_0}_{2^\ekk}$, $\cdots$, $\share{\Vec{a}_{m-1}}_{2^\ekk}$ and a secret-shared label vector $\share{\Vec{y}}_{2^\ekk}$ for sharing. That is, for the $i$-th sample, its attributes and label are $\Vec{a}_0[i]$, $\cdots$, $\Vec{a}_{m-1}[i]$, and $\Vec{y}[i]$.

 Besides, we assign each node a node ID ($\mathit{nid}$ for short) to reflect the relationship among the nodes. That is, for the node whose $\mathit{nid}$ is $j$ in the $k$-th layer, its left child node is the node whose $\mathit{nid}$ is $2*j + 1$ in the ($k+1$)-th layer. Its right child node is the node whose $\mathit{nid}$ is $2*j + 2$ in the ($k+1$)-th layer. 

Furthermore, we leverage a secret-shared sample-node vector   $\share{\vv{\mathit{spnd}}^{(k)}}_{2^\ekk}$  to privately indicate which node the samples belong to in the $k$-th layer of a decision tree. Concretely, the equation $\vv{\mathit{spnd}}^{(k)}[i] = j$ indicates the $i$-th sample belongs to the node whose $nid$ is $j$ in the $k$-th layer.




\section{Design of \fname}
 
\subsection{Protocol for Training a Decision Tree}

As is shown in Protocol~\ref{pro:traindecisiontree}, the protocol \textit{TrainDecisionTree} is used to securely train a decision tree.  The parties in this protocol input $m$ secret-shared attribute vectors $\share{\vec{a}_0}_{2^\ekk},$ $\cdots,$ $\share{\vec{a}_{m-1}}_{2^\ekk}$, a secret-shared label vector $\share{\vec{y}}_{2^\ekk}$, and an integer $h$ (representing the desired tree's height). They get $h+1$ secret-shared layers $\share{layer^{(0)}}_{2^\ekk}$ ,$ \cdots$, $\share{layer^{(h)}}_{2^\ekk}$ of a trained decision tree as the output. If the $k$-th layer is an internal layer,  the secret-shared layer $\share{layer^{(k)}}_{2^\ekk}$ contains the secret-shared $\mathit{nid}$, split attribute index, and split threshold of each node in the $k$-th layer. If the $k$-th layer is a leaf layer,  the secret-shared layer $\share{layer^{(k)}}_{2^\ekk}$ contains the secret-shared $\mathit{nid}$ and predicted label of each node in the $k$-th layer. 

\begin{algorithm}
    \LinesNumbered
    \small
    \caption{$\mathit{TrainDecisionTree}$}
    \label{pro:traindecisiontree}
    \begin{flushleft}
     \textbf{Input:}  $m$ secret-shared attribute vectors, $\share{\vec{a}_0}_{2^\ekk},\cdots,\share{\vec{a}_{m-1}}_{2^\ekk}$, a secret-shared label vector $\share{\vec{y}}_{2^\ekk}$, and an  integet $h$.  \\
     \textbf{Output:}  $h+1$ secret-shared layers,$\share{layer^{(0)}}_{2^\ekk}$, $\cdots$,$ \share{layer^{(h)}}_{2^\ekk}$, of a  decision tree.  \\
    \end{flushleft}
    \begin{algorithmic}[1]
 \FOR{ each $ i \in [0, m-1]$ in parallel }
     \STATE $\share{\vec{\pi}_i}_{2^\ekk} = \mathit{GenPerm}  $ $(\share{\vec{a}_i}_{2^\ekk})$.
\ENDFOR
    \STATE $\share{\vv{\mathit{spnd}}^{(0)}[j]}_{2^\ekk} = \share{0}$ for each $j \in [0, n-1]$.
    \FOR{ each $ k \in [0, h-1]$  }
        \STATE $ \share{layer^{(k)}}_{2^\ekk}$, $\share{\vv{\mathit{spat}}^{(k)}}_{2^\ekk}$, $\share{\vv{\mathit{spth}}^{(k)}}_{2^\ekk}$=$ \mathit{TrainInternalLayer}$( $k$, $ \share{\vv{\mathit{spnd}}^{(k)}}_{2^\ekk}$, $\share{\vec{a}_0}_{2^\ekk}$,$\cdots,\share{\vec{a}_{m-1}}_{2^\ekk}$, $ \share{\vec{y}}_{2^\ekk}$, $ \share{\vec{\pi}_0}_{2^\ekk}$,$\cdots$, $\share{\vec{\pi}_{m-1}}_{2^\ekk})$.
        \STATE $\share{\vec{b}}_{2^\ekk} $=$ \mathit{TestSamples}(\share{\vec{a}_0}_{2^\ekk}$,$\cdots,\share{\vec{a}_{m-1}}_{2^\ekk}$,$ \share{\vv{\mathit{spat}}^{(k)}}_{2^\ekk}$, $\share{\vv{\mathit{spth}}^{(k)}}_{2^\ekk})$.
       \STATE $\share{\vv{\mathit{spnd}}^{(k+1)}}_{2^\ekk} = 2* \share{\vv{\mathit{spnd}}^{(k)}}_{2^\ekk} + 1 + \share{\vec{b}}_{2^\ekk}$.
    \STATE $\share{\vec{\pi}_0}_{2^\ekk},\cdots,\share{\vec{\pi}_{m-1}}_{2^\ekk} = \mathit{UpdatePerms}$ $(\share{\vec{b}}_{2^\ekk}$, $\share{\vec{\pi}_0}_{2^\ekk}$,$\cdots$,  $\share{\vec{\pi}_{m-1}}_{2^\ekk})$.
    \ENDFOR
    \STATE $\share{\mathit{layer}^{(h)}}_{2^\ekk} = \mathit{TrainLeafLayer}(h, \share{\vec{\pi}_0}_{2^\ekk}, \share{\vv{\mathit{spnd}}^{(h)}}_{2^\ekk}, \share{\vec{y}}_{2^\ekk})$.

 \end{algorithmic}
\end{algorithm}

The protocol \textit{TrainDecisionTree} consists of three stages. (1) The first stage (Line 1-4): the parties generate $m$ secret-shared permutations from the attributes by calling the protocol \textit{GenPerm} (Protocol~\ref{pro:genperm}) and initial a secret-shared sample-node vector $\share{\vv{\mathit{spnd}}^{(0)}}_{2^\ekk}$. These permutations are used to securely sort the training samples to save secure comparison operations. The parties set all elements of $\share{\vv{\mathit{spnd}}^{(0)}}_{2^\ekk}$ to $\share{0}$, since all samples belong to the root node whose $\mathit{nid}$ is 0 in the 0-th layer. (2) The second stage (Line 5-10): the parties securely train each internal layer of the decision tree, compute the secret-shared sample-node vector for the next layer, and update the pre-generated permutations. To securely train an internal layer, the parties call the protocol \textit{TrainInternalLayer} (Protocol~\ref{pro:traininternallayer}), which outputs a secret-shared layer $\share{layer^{(k)}}_{2^\ekk}$, a secret-shared sample-attribute vector $\share{\vv{\mathit{spat}}^{(k)}}_{2^\ekk}$, and a secret-shared sample-threshold vector $\share{\vv{\mathit{spth}}^{(k)}}_{2^\ekk}$ (Line 6). $\vv{\mathit{spat}}^{(k)}$ and $\vv{\mathit{spth}}^{(k)}$ contain the split attribute index and split threshold of the node to which the $i$-th sample belongs in the $k$-th layer, respectively. For example, the equations $\vv{\mathit{spat}}^{(2)}[0]=1$ and $\vv{\mathit{spth}}^{(2)}[0]=5$ represents that the split attribute index and split threshold of the node to which the $0$-th sample belongs in the $2$-th layer are $1$ and $5$, respectively.  The parties then obtain the secret-shared comparison results $\share{\vec{b}}_{2^\ekk}$ between the samples and the thresholds by calling the protocol \textit{TestSamples} (Protocol \ref{pro:testsamples} in Appendix~\ref{appendix.pro}) (Line 7). These secret-shared comparison results are used to compute the secret-shared vector $\share{\vv{\mathit{spnd}}^{(k+1)}}_{2^\ekk}$, and update the pre-generated secret-shared permutations $\share{\vec{\pi}_0}_{2^\ekk},$ $\cdots,$ $\share{\vec{\pi}_{m-1}}_{2^\ekk}$ (Line 8-9).  The idea for computing  $\share{\vv{\mathit{spnd}}^{(k+1)}}_{2^\ekk}$ is as follows:  if a sample in the $k$-th layer belongs to a node whose $\mathit{nid}$ is $j$ and has a comparison result of `0', it will be assigned to the left child node ($\mathit{nid}$ is $2*j + 1$) in the $(k+1)$-th layer. Conversely, if the result is `1', the sample is assigned to the right child node ($\mathit{nid}$ is $2*j + 2$) in the $(k+1)$-th layer. To update the pre-generated permutations, the parties call the protocol \textit{UpdatePerms} (Protocol~\ref{pro:updateperms}).  (3) The third stage (Line 11):  the parties securely train the leaf layer by calling the protocol \textit{TrainLeafLayer} 
 (Protocol \ref{pro:trainleaflayer} in Appendix~\ref{appendix.pro}).

To optimize the communication overhead for securely training a decision tree, we propose two communication optimizations: One is based on the secure radix sort protocols~\cite{chida2019efficient} and mainly implemented in our proposed protocols \textit{UpdatePerms} (Protocol~\ref{pro:updateperms} in Section~\ref{sec.update_permuation}),  \textit{TrainInternalLayer} (Protocol~\ref{pro:traininternallayer} in Section~\ref{sec.train_internal_layer}), and \textit{TrainLeafLayer} (Protocol~\ref{pro:trainleaflayer} in Appendix~\ref{appendix.pro}). The other is based on a novel share conversion protocol proposed by us and mainly implemented in our proposed protocols \textit{ComputeModifiedGini} (Protocol~\ref{pro:computemodifiedgini} in Section~\ref{sec.gini_protocol})  and \textit{ConvertShare} (Protocol~\ref{pro:convertshare} in Section~\ref{sec.shareconversion}).  

Note that we present an example to show the key steps of the training process in Figure~\ref{fig.trainingprocess} in Appendix~\ref{app.example} for understanding.

\subsection{Communication Optimization Based on Secure Radix Sort Protocols}~\label{sec.rsp}


\subsubsection{Brief Overview of Existing Methods for Secure Dataset Splitting}
\label{sec.briefoverview}

Securely splitting a dataset with continuous attributes usually requires numerous secure comparison operations, leading to huge communication overhead. To resolve this issue, Abspoel et al.\cite{MarkAbspoel2021SecureTO} propose securely generating secret-shared permutations from the secret-shared continuous attribute vectors at the beginning of the training process. These pre-generated secret-shared permutations are used to securely sort the training samples according to the attribute vectors, so that lots of secure comparison operations can be saved. However, to hide the splitting information of the training dataset, their method requires securely training each node with a padded dataset of the same size as the entire training dataset. The secure training process with padded datasets results in huge communication overhead that grows exponentially as the height of the decision tree increases.

To achieve linear growth in communication overhead as the height of the decision tree increases, Hamada et al.~\cite{hamada2021efficient} propose a series of group-wise protocols. These protocols treat the samples of each node as a group and enable securely training all nodes of one layer simultaneously with non-padded datasets, thereby avoiding the exponential growth in communication overhead. However, the group-wise protocols are incompatible with pre-generated secret-shared permutations. Specifically, to use the group-wise protocols, the samples belonging to the same node must be stored consecutively. Nevertheless, after using the pre-generated secret-shared permutations to securely sort the samples, the samples belonging to the same node will no longer be stored consecutively.
 To resolve this issue, Hamada et al.~\cite{hamada2021efficient} propose securely sorting the training samples first according to the attribute vectors and then according to the sample-node vector in each layer. Due to the stability of the sort, this method can consecutively store samples belonging to the same node and securely sort the samples according to the attribute vectors within each node. Nevertheless, this method requires repeatedly generating secret-shared permutations from each attribute and the sample-node vector for training each internal layer of the decision tree. The repeated generation process also results in significant communication overhead.

\subsubsection{Main Idea}
To optimize the communication overhead, we employ the secure radix sort protocols~\cite{chida2019efficient} to securely update the pre-generated secret-shared permutations, so that the pre-generated secret-shared permutations become compatible with the group-wise protocols after the updating process. More specifically, once updated, the pre-generated secret-shared permutations can be employed to consecutively store samples belonging to the same node, and securely sort these samples according to their attribute vectors within each node. This method combines the advantages of the methods proposed by Abspoel et al.~\cite{MarkAbspoel2021SecureTO} and Hamada et al.~\cite{hamada2021efficient}: this method only requires securely generating the secret-shared permutations once, and the communication overhead of this method for training a decision tree only grows linearly as the height of the decision tree increases.

The principle behind securely updating the pre-generated secret-shared permutations based on the secure radix sort protocols~\cite{chida2019efficient} is as follows: the comparison result between a sample and the split threshold of a node $\mathcal{N}$ that the sample belongs to determines which node the sample should be assigned to. A comparison result of `0' indicates this sample should be assigned to the left child node of $\mathcal{N}$, while a comparison result of `1' indicates this sample should be assigned to the right child node of $\mathcal{N}$. Thus, if the samples are sorted according to the comparison result vector, the samples belonging to the same node will be stored consecutively. Furthermore, if the pre-generated permutations are updated with the comparison result vector, due to the stability of radix sort, the updated permutations can then be used to store the samples belonging to the same node consecutively, and meanwhile, sort the samples according to their attribute vectors within each node. 


\subsubsection{Protocol for Updating Permutations}
\label{sec.update_permuation}

\begin{algorithm}
    \LinesNumbered
    \small
    \caption{$\mathit{UpdatePerms}$}
    \label{pro:updateperms}
    \begin{flushleft}
     \textbf{Input:} A secret-shared comparison result vector $\share{\vec{b}}_{2^\ekk}$ and  $m$ secret-shared permutations $\share{\vec{\pi}_0}_{2^\ekk},\cdots,\share{\vec{\pi}_{m-1}}_{2^\ekk}$.  \\
 \textbf{Output:}  $m$ secret-shared  permutations $\share{\vec{\pi}_0}_{2^\ekk},\cdots,\share{\vec{\pi}_{m-1}}_{2^\ekk}$. \\
 \end{flushleft}

 \begin{algorithmic}[1]
    \FOR{ each $ i \in [0, m-1]$ in parallel }
        \STATE $ \share{\vec{b'}_i}_{2^\ekk} = \mathit{ApplyPerm}(\share{\pi_i}_{2^\ekk}, \share{\vec{b}}_{2^\ekk})$.
        \STATE $ \share{\vec{\alpha}}_{2^\ekk} =  \mathit{GenPermByBit}(\share{\vec{b'}_i}_{2^\ekk})$.
        \STATE $ \share{\vec{\pi}_i}_{2^\ekk} =  \mathit{ComposePerms}(\share{\vec{\pi}_i}_{2^\ekk}, \share{\vec{\alpha}}_{2^\ekk})$.
    \ENDFOR
 \end{algorithmic}
\end{algorithm}

As is shown in Protocol~\ref{pro:updateperms}, the protocol \textit{UpdatePerms} securely updates the pre-generated permutations with the comparison result vector that contains the comparison results between the samples and the thresholds. The parties in this protocol input a secret-shared comparison result vector $\share{\vec{b}}_{2^\ekk}$ and $m$ secret-shared permutations $\share{\vec{\pi}_0}_{2^\ekk},\cdots,\share{\vec{\pi}_{m-1}}_{2^\ekk}$. They get $m$ updated secret-shared permutations as the output. 

The protocol \textit{UpdatePerms} follows the updating step (Line 4-6) of the protocol \textit{GenPerm} (Protocol~\ref{pro:genperm}) to update each secret-shared permutations $\share{\vec{\pi}_i}_{2^\ekk}$:  the parties first apply $\share{\vec{\pi}_i}_{2^\ekk}$ to  $\share{\vec{b}}_{2^\ekk}$ to generate a new secret-shared vector $\share{\vec{b'}_i}_{2^\ekk}$ (Line 2). Next, the parties generate a secret-shared permutation $\share{\vec{\alpha}}_{2^\ekk}$ from  $\share{\vec{b'}_i}_{2^\ekk}$ by calling the protocol \textit{GenPermByBit} (introduced in Section~\ref{sec.sort}) (Line 3), such that $\share{\vec{\alpha}}_{2^\ekk}$ can be used to sort a secret-shared vector that has been sorted with $\share{\vec{\pi}_i}_{2^\ekk}$ further according to $\share{\vec{b}}_{2^\ekk}$. Finally, the parties compose the effect of $\share{\vec{\pi}_i}_{2^\ekk}$ and $\share{\vec{\alpha}}_{2^\ekk}$ by calling the protocol \textit{ComposePerms} (introduced in Section~\ref{sec.sort}), resulting in an updated $\share{\vec{\pi}_i}_{2^\ekk}$ (Line 4).

\subsubsection{Protocol for Training Internal Layer.}

\label{sec.train_internal_layer}

As is shown in Protocol~\ref{pro:traininternallayer}, the protocol \textit{TrainInternalLayer} is used to securely train an internal layer of a decision tree with the secret-shared permutations.  The parties in this protocol input an integer $k$, representing the height of the internal layer, a secret-shared sample-node vector $\share{\vv{\mathit{spnd}}}_{2^\ekk}$, $m$ secret-shared attribute vectors $\share{\vec{a}_0}_{2^\ekk},\cdots,\share{\vec{a}_{m-1}}_{2^\ekk}$, a secret-shared label vector $\share{\vec{y}}_{2^\ekk}$, and $m$ secret-shared permutations $\share{\vec{\pi}_0}_{2^\ekk},\cdots,\share{\vec{\pi}_{m-1}}_{2^\ekk}$. They get a secret-shared layer $\share{\mathit{layer}}_{2^\ekk}$, a secret-shared sample-attribute vector $\share{\vv{\mathit{spat}}}_{2^\ekk}$, and a secret-shared sample-threshold vector $\share{\vv{\mathit{spth}}}_{2^\ekk}$ as the output.

\begin{algorithm}
    \LinesNumbered
    \small
    \caption{$\mathit{TrainInternalLayer}$}
    \label{pro:traininternallayer}
    \begin{flushleft}
  \textbf{Input:} An integer $k$ that represents the height of the internal layer,  a secret-shared sample-node vector $\share{\vv{\mathit{spnd}}}_{2^\ekk}$, $m$ secret-shared attribute vectors, $\share{\vec{a}_0}_{2^\ekk},\cdots,\share{\vec{a}_{m-1}}_{2^\ekk}$, a secret-shared label vector $\share{\vec{y}}_{2^\ekk}$,  and  $m$ secret-shared permutations $\share{\vec{\pi}_0}_{2^\ekk},\cdots,\share{\vec{\pi}_{m-1}}_{2^\ekk}$.  \\
 \textbf{Output:}  a secret-shared layer $\share{\mathit{layer}}_{2^\ekk}$, a secret-shared sample-attribute vector $\share{\vv{\mathit{spat}}}_{2^\ekk}$, and a secret-shared sample-threshold vector $\share{\vv{\mathit{spth}}}_{2^\ekk}$.  \\
 \end{flushleft}

 \begin{algorithmic}[1]
    \STATE $ \share{\vv{\mathit{spnd'}}}_{2^\ekk} = \mathit{ApplyPerm}(\share{\vec{\pi}_0}_{2^\ekk}, \share{\vv{\mathit{spnd}}}_{2^\ekk})$.
    \STATE $\share{\vec{g}[0]}_{2^\ekk} = \share{1}_{2^\ekk}.$ 
    \STATE $\share{\vec{g}[j]}_{2^\ekk} = (\share{\vv{\mathit{spnd'}}[j-1]}_{2^\ekk} \neq \share{\vv{\mathit{spnd'}}[j]}_{2^\ekk})$ for each $j \in [1, n-1]$.
    \FOR{ each $ i \in [0, m-1]$ in parallel }
        \STATE $ \share{\vec{y'}}_{2^\ekk} = \mathit{ApplyPerm}(\share{\vec{\pi}_i}_{2^\ekk}, \share{\vec{y}}_{2^\ekk})$.
       \STATE $ \share{\vec{a'}_i}_{2^\ekk} = \mathit{ApplyPerm}(\share{\vec{\pi}_i}_{2^\ekk}, \share{\vec{a}_i}_{2^\ekk})$.
      \STATE $ 
      \share{\vv{\mathit{spth}}_i}_{2^\ekk}, \share{\vv{\mathit{gini}}_i}_{2^\ekk}  = \mathit{AttributeWiseSplitSelection}$ \\ $(\share{\vec{g}}_{2^\ekk}, \share{\vec{a'}_i}_{2^\ekk}, \share{\vec{y'}}_{2^\ekk})$.
    \ENDFOR
    \FOR{ each $ j \in [0, n-1]$ in parallel }
        \STATE $ \share{\vv{\mathit{ginilist}}}_{2^\ekk} = [\share{\vv{\mathit{gini}}_0[j]}_{2^\ekk}, \cdots,$
         $  \share{\vv{\mathit{gini}}_{m-1}[j]}_{2^\ekk}]$.
        \STATE $ \share{\vv{\mathit{aidxlist}}}_{2^\ekk} = [\share{0}_{2^\ekk}, \cdots,$
         $  \share{m-1}_{2^\ekk}]$.
        \STATE $ \share{\vv{\mathit{stlist}}}_{2^\ekk} = [\share{\vv{\mathit{spth}}_0[j]}_{2^\ekk}, \cdots, \share{\vv{\mathit{spth}}_{m-1}[j]}_{2^\ekk}]$.
        \STATE $ \share{\vv{\mathit{spat}}[j]}_{2^\ekk} = \mathit{VectMax}(\share{\vv{\mathit{ginilist}}}_{2^\ekk}, \share{\vv{\mathit{aidxlist}}}_{2^\ekk})$.
        \STATE $ \share{\vv{\mathit{spth}}[j]}_{2^\ekk} = \mathit{VectMax}(\share{\vv{\mathit{ginilist}}}_{2^\ekk}, \share{\vv{\mathit{stlist}}}_{2^\ekk}).$
    \ENDFOR
    \STATE $\share{\mathit{layer}}_{2^\ekk}$ = $\mathit{FormatLayer}$ $(k, $ $\share{\vec{g}}_{2^\ekk},$  $\share{\vv{\mathit{spnd'}}}_{2^\ekk},$ $ \share{\vv{\mathit{spat}}}_{2^\ekk}, $ $\share{\vv{\mathit{spth}}}_{2^\ekk})$.

    \STATE $\share{\vv{\mathit{spat}}}_{2^\ekk} = \mathit{UnApplyPerm}(\share{\vec{\pi}_0}_{2^\ekk}, \share{\vv{\mathit{spat}}}_{2^\ekk})$.
    \STATE $\share{\vv{\mathit{spth}}}_{2^\ekk} = \mathit{UnApplyPerm}(\share{\vec{\pi}_0}_{2^\ekk}, \share{\vv{\mathit{spth}}}_{2^\ekk})$.

 \end{algorithmic}
\end{algorithm}
 
The protocol \textit{TrainInternalLayer} consists of four stages. (1) The first stage (Line 1-3): the parties compute a secret-shared group flag vector $\share{\vec{g}}_{2^\ekk}$ that privately indicates the boundaries of the nodes. To compute $\share{\vec{g}}_{2^\ekk}$, the parties first apply one of the permutations (without loss of generality, $\share{\vec{\pi}_0}_{2^\ekk}$ is chosen) to $\share{\vv{\mathit{spnd}}}_{2^\ekk}$ to obtain a new secret-shared sample-node vector $\share{\vv{\mathit{spnd'}}}_{2^\ekk}$ (Line 1). This process ensures that the elements with the same values in $\share{\vv{\mathit{spnd'}}}_{2^\ekk}$ are stored consecutively. Thus, if $\vv{\mathit{spnd'}}[j] \neq \vv{\mathit{spnd'}}[j-1]$, the $j$-th sample should be the first sample of a node. Then, the parties set $\share{\vec{g}[0]}_{2^\ekk}$ to $\share{1}$  and compute the remaining elements of $\share{\vec{g}}_{2^\ekk}$ based on $\share{\vv{\mathit{spnd'}}}_{2^\ekk}$ (Line 2-3). (2) The second stage (Line 4-8): the parties securely compute the split thresholds with each attribute and the corresponding modified Gini impurities. To achieve this, the parties first apply each secret-shared permutation $\share{\vec{\pi}_i}_{2^\ekk}$ to $\share{\vec{y}}_{2^\ekk}$ and $\share{\vec{a}_i}_{2^\ekk}$ to generate a new secret-shared label vector $\share{\vec{y'}}_{2^\ekk}$ and a new secret-shared attribute vector $\share{\vec{a'}_i}_{2^\ekk}$ (Line 5-6). This process makes the elements of $\share{\vec{y'}}_{2^\ekk}$ and $\share{\vec{a'}_i}_{2^\ekk}$ belonging to the same node stored consecutively and sorted according to the corresponding attribute vector within each node, such that the group-wise protocols~\cite{hamada2021efficient} can be applied on  $\share{\vec{y'}}_{2^\ekk}$ and $\share{\vec{a'}_i}_{2^\ekk}$. Subsequently, the parties call the protocol \textit{AttributeWiseSplitSelection} (Protocol~\ref{pro:attributewisesplitselection} in Appendix~\ref{appendix.pro}) to obtain a secret-shared sample-threshold vector $\share{\vv{\mathit{spth}}_i}_{2^\ekk}$ and a secret-shared Gini impurity vector $\share{\vv{\mathit{gini}}_i}_{2^\ekk}$ (Line 7). $\vv{\mathit{spth}}_i[j]$ is the split threshold of the node to which the $j$-sample belongs and is computed with only the $i$-th attribute.  $\vv{\mathit{gini}}_i[j]$ is the modified Gini impurity for the split point $(a_i, \vv{\mathit{spth}}_i[j])$. (3) The third stage (Line 9-15): the parties securely find the split attribute index and the split threshold with a maximum modified Gini impurity across different attributes. This is achieved by calling the protocol \textit{VectMax} (introduced in Section~\ref{sec.vectmax}). (4) The fourth stage (Line 16-18): the parties first compute the secret-shared layer $\share{\mathit{layer}}_{2^\ekk}$ by calling the protocol \textit{FormatLayer} (Protocol~\ref{pro:formatlayer} in Appendix~\ref{appendix.pro}) (Line 16). The protocol \textit{FormatLayer} is used to remove the redundant values from the input vectors to leave only one $\mathit{nid}$, split attribute index, and split threshold for each node. Then the parties reverse the effect of $\share{\vec{\pi}_0}_{2^\ekk}$ on the secret-shared sample-attribute vector $\share{\vv{\mathit{spat}}}_{2^\ekk}$ and sample-threshold vector $\share{\vv{\mathit{spth}}}_{2^\ekk}$  to align their element positions with the original secret-shared attribute vectors (Line 17-18). So that,  $\share{\vv{\mathit{spat}}}_{2^\ekk}$ and  $\share{\vv{\mathit{spth}}}_{2^\ekk}$ can be used in the comparison between the samples and the split thresholds (Line 7 of Protocol~\ref{pro:traindecisiontree}).

\subsection{Communication Optimization Based on Share Conversion Protocol}
\label{sec.sep}

\subsubsection{Brief Overview of Existing Methods for Secure Splitting Criterion Computation.}
The existing multi-party training frameworks~\cite{hamada2021efficient, MarkAbspoel2021SecureTO, deHoogh2014Practical} for decision trees perform almost all computations on a large ring to accommodate the secure computations for the splitting criterion, e.g. the modified Gini impurity. The secure computations for the splitting criterion usually involve several sequential 
bit-length expansion operations, such as secure multiplication and division operations.  These operations usually produce intermediate or final results that require twice as many bits for representation compared to their inputs.   For instance, a secure multiplication operation with two 10-bit numbers as input produces a 20-bit result. Performing such operations sequentially several times requires a large ring to represent the intermediate or final results. For example,  the framework proposed by Abspoel et al.~\cite{MarkAbspoel2021SecureTO}  requires a ring with bit length at least $5 \lceil \log \frac{n}{2} \rceil$ to accommodate the secure computation for the modified Gini impurity, where $n$ is the number of samples in the training dataset. Performing almost all computations on a large ring incurs huge communication overhead to the existing multi-party training frameworks.

\subsubsection{Main Idea}
\label{sec.gini_protocol}
To optimize the communication overhead incurred by performing almost all computations on a large ring, we propose an efficient share conversion protocol
to  convert shares between a small ring and a large ring.   We observe that excluding the secure computations for the splitting criterion,  other computations in the training process only involve bit-length invariant operations, such as secure addition and comparison operations, which cause at most a single-bit increase and can be performed on a small ring whose bit length is just two bits large than $\lceil \log n \rceil$ (one additional bit for representing the sign and the other additional bit for avoiding overflow when performing secure addition operations). With our proposed share conversion protocol, the parties in \fname can  convert shares to a large ring when performing the bit-length extension operations and then  convert the shares back to a small ring once these operations are completed. As a result,  the communication overhead of the bit-length invariant operations can be significantly reduced.

\subsubsection{Protocol for Computing Modified Gini Impurity} As is shown in Protocol~\ref{pro:computemodifiedgini},
   the protocol \textit{ComputeModifiedGini} is based on our proposed share conversion protocol \textit{ConvertShare} (Protocol~\ref{pro:convertshare}) to convert shares to a large ring $\mathbb{Z}_{2^\ell}$ for accommodating the bit-length expansion operations. Besides, the computed modified Gini impurity vector is finally converted back to a small ring $\mathbb{Z}_{2^\ekk}$ to save the communication overhead of the further bit-length invariant operations.  The parties in this protocol input a secret-shared group flag vector $\share{\vec{g}}_{2^\ekk}$ and a secret-shared label vector $\share{\vec{y}}_{2^\ekk}$, whose elements belonging to the same nodes are stored consecutively and sorted according to an attribute vector $\vec{a}_i$ within each node.  The parties get a secret-shared modified Gini impurity vector $\share{\vv{\mathit{gini}}}_{2^\ekk}$ as the output, where $\vv{\mathit{gini}}[j]$ is the modified Gini impurity of the split point ($a_i, t_j$), where $t_j = (\vec{a}_i[j] + \vec{a}_i[j+1])/2$ for each $j \in [0, n-2]$ and $t_{n-1} = \vec{a}_i[n-1]$.

\begin{algorithm}
    \LinesNumbered
    \small
    \caption{$\mathit{ComputeModifiedGini}$}
    \label{pro:computemodifiedgini}
    \begin{flushleft}
  \textbf{Input:}   A secret-shared group flag vector $\share{\vec{g}}_{2^\ekk}$, and a secret-shared label vectors $\share{\vec{y}}_{2^\ekk}$. \\
 \textbf{Output:} A secret-shared modified Gini impurity vector $\share{\vv{\mathit{gini}}}_{2^\ekk}$. \\
 \end{flushleft}

 \begin{algorithmic}[1]
     \STATE $ \share{\vv{\mathit{one}}[j]}_{2^\ekk} = \share{1}_{2^\ekk} $ for each $ j \in [0, n-1]$.
    \STATE $\share{\vv{\mathit{ts}}}_{2^\ekk} = \mathit{GroupSum}(\share{\vec{g}}_{2^\ekk}, \share{\vv{\mathit{one}}}_{2^\ekk}) $.
    \STATE $\share{\vv{\mathit{ps}}}_{2^\ekk} = \mathit{GroupPrefixSum}(\share{\vec{g}}_{2^\ekk}, \share{\vv{\mathit{one}}}_{2^\ekk}) $.
    \STATE $\share{\vv{\mathit{ss}}}_{2^\ekk} = \share{\vv{\mathit{ts}}}_{2^\ekk} - \share{\vv{\mathit{ps}}}_{2^\ekk} $.
    \STATE $\share{\vv{\mathit{ps}}}_{2^\ell} = \mathit{ConvertShare}(\share{\vv{\mathit{ps}}}_{2^\ekk}, \ekk, \ell) $.
    \STATE $\share{\vv{\mathit{ss}}}_{2^\ell} = \mathit{ConvertShare}(\share{\vv{\mathit{ss}}}_{2^\ekk}, \ekk, \ell) $.
    \FOR{ each $ l \in [0, v-1]$ in parallel } 
        \STATE $ \share{\vec{y}_l}_{2^\ekk} =  \share{\vec{y}}_{2^\ekk} \overset{?}{=} l $.
        \STATE $ \share{\vec{c}_l}_{2^\ekk} = \mathit{GroupSum}(\share{\vec{g}}_{2^\ekk}, \share{\vec{y}_l}_{2^\ekk}) $.
       \STATE $ \share{\vv{\mathit{pre}}_l}_{2^\ekk} = \mathit{GroupPrefixSum}(\share{\vec{g}}_{2^\ekk}, \share{\vec{y}_l}_{2^\ekk}) $.
        \STATE $ \share{\vv{\mathit{suf}}_l}_{2^\ekk} = \share{\vec{c}_l}_{2^\ekk} - \share{\vv{\mathit{pre}}_l}_{2^\ekk}$.
       \STATE $ \share{\vv{\mathit{pre}}_l}_{2^\ell} = \mathit{ConvertShare}(\share{\vv{\mathit{pre}}_l}_{2^\ekk}, \ekk, \ell)$. 
      \STATE $ \share{\vv{\mathit{suf}}_l}_{2^\ell} = \mathit{ConvertShare}(\share{\vv{\mathit{suf}}_l}_{2^\ekk}, \ekk, \ell)$. 
       \STATE $ \share{\vv{\mathit{presq}}_l}_{2^\ell} = \share{\vv{\mathit{pre}}_l}_{2^\ell} * \share{\vv{\mathit{pre}}_l}_{2^\ell}$.
       \STATE $ \share{\vv{\mathit{sufsq}}_l}_{2^\ell} = \share{\vv{\mathit{suf}}_l}_{2^\ell} * \share{\vv{\mathit{suf}}_l}_{2^\ell}$.
    \ENDFOR
    \STATE $ \share{\vv{\mathit{presqs}}}_{2^\ell} = \sum_{l=0}^{v-1}\share{\vv{\mathit{presq}}_l}_{2^\ell} $.
    \STATE $ \share{\vv{\mathit{sufsqs}}}_{2^\ell} = \sum_{l=0}^{v-1}\share{\vv{\mathit{sufsq}}_l}_{2^\ell} $.
    \STATE $\share{\vv{\mathit{gini}}}_{2^\ell} = \share{\vv{\mathit{presqs}}}_{2^\ell} / \share{\vv{\mathit{ps}}}_{2^\ell} $  $+ \share{\vv{\mathit{sufsqs}}}_{2^\ell} / \share{\vv{\mathit{ss}}}_{2^\ell} $.
    \IF{$f + \lceil \log n \rceil > \ekk -1$}
    \STATE $\share{\vv{\mathit{gini}}}_{2^\ell} = \mathit{Trunc}(\share{\vv{\mathit{gini}}}_{2^\ell},f + \lceil \log n \rceil -( \ekk -1))$.
    \ENDIF
    \STATE $\share{\vv{\mathit{gini}}}_{2^\ekk} = \mathit{ConvertShare}(\share{\vv{\mathit{gini}}}_{2^\ell}, \ell, \ekk)$.
\end{algorithmic}
\end{algorithm}

 The protocol \textit{ComputeModifiedGini} consists of three stages. (1) The first stage (Line 1-6): the parties securely compute the values of $|\mathcal{D}^{node}_{a_i < t_j}|$ and $|\mathcal{D}^{node}_{a_i \geq t_j}|$ for each split point ($a_i, t_j$) simultaneously.  To do this, the parties first create a secret-shared one vector $\share{\vv{\mathit{one}}}_{2^\ekk}$ (Line 1), and compute the secret-shared group-wise prefix sum $\share{\vv{ps}}_{2^\ekk}$ and group-wise suffix sum $\share{\vv{\mathit{ss}}}_{2^\ekk}$ of $\share{\vv{\mathit{one}}}_{2^\ekk}$(Line 2-4).   $\vv{\mathit{ps}}$ and  $\vv{\mathit{ss}}$ contain all the values of $|\mathcal{D}^{node}_{a_i < t_j}|$ and $|\mathcal{D}^{node}_{a_i \geq t_j}|$, because the samples have been sorted according to the attribute vector $\vec{a}_i$. The shares of $\vv{\mathit{ps}}$ and  $\vv{\mathit{ss}}$ are then converted to a large ring $\mathbb{Z}_{2^\ell}$ (Line 5-6) to accommodate the secure division operation. (2) The second stage (Line 7-18): the parties securely compute the values of $\sum_{l=0}^{v-1}|\mathcal{D}^{node}_{a_i < t_j  \wedge y=l}|^2$ and $\sum_{l=0}^{v-1}|\mathcal{D}^{node}_{a_i \geq t_j  \wedge y=l}|^2$ for each split point ($a_i, t_j$) simultaneously. To do this,  the parties first compute a secret-shared vector $\share{\vec{y}_l}_{2^\ekk}$ that privately indicates whether the samples' labels are $l$ (Line 8). The parties then  compute the corresponding secret-shared group-wise prefix sum $\share{\vv{\mathit{pre}}_l}_{2^\ekk}$ and group-wise suffix sum  $\share{\vv{\mathit{suf}}_l}_{2^\ekk}$ for the label $l$ (Line 9-11). The shares of these values are then converted to the large ring $\mathbb{Z}_{2^\ell}$ for computing their squared values $ \share{\vv{\mathit{presq}}_l}_{2^\ell}$ and $\share{\vv{\mathit{sufsq}}_l}_{2^\ell}$ (Line 12-15). Subsequently, the parties  sum up the squared values $ \share{\vv{\mathit{presq}}_l}_{2^\ell}$ and $\share{\vv{\mathit{sufsq}}_l}_{2^\ell}$ of each label (Line 17-18). (3) The third stage (Line 19-23): the parties first compute the secret-share modified Gini impurity vector $\share{\vv{\mathit{gini}}}_{2^\ell}$,  then truncate $\share{\vv{\mathit{gini}}}_{2^\ell}$ to leave the most significant $\ekk-1$ bits, and finally convert the truncated results to a small ring $\mathbb{Z}_{2^\ekk}$. Note that before truncation, a modified Gini impurity is a value in $[1,n)$  and has at most $ \lceil \log n \rceil + f$ bits, where $f$ is the bit length to represent the decimal part. Thus, only if $ \lceil \log n \rceil + f > \ekk -1$, does the truncation operation need to be performed.

\subsubsection{Design of Share Conversion Protocol}
\label{sec.shareconversion}
To efficiently and accurately perform the protocol \textit{ComputeModifiedGini}, a share conversion protocol that is both efficient and correct is necessary. Though several share conversion protocols have been proposed~\cite{kelkar2022secure, aly2019zaphod, liu2020mpc, gupta2022llama, jawalkar2023orca}, these existing protocols either lack efficiency or have a high failure probability. Therefore, in this section, we present an efficient share conversion protocol, \textit{ConvertShare}, which only requires an online communication size of $4\ell + 4$ bits in a single communication round and ensures correctness. 


Note that converting the shares of a secret $x$ ($x \in [0,2^{\ekk-1})$) from a large ring $\mathbb{Z}_{2^\ell}$ to a small ring $\mathbb{Z}_{2^\ekk}$ can be achieved just by a local modular operation~\cite{rathee2021sirnn}~\footnote{Converting the shares of a secret $x$ ($x \in [0,2^{\ekk-1}$) from a large ring $\mathbb{Z}_{2^\ell}$ to a small ring $\mathbb{Z}_{2^\ekk}$ is referred to as `Reduce' in the study~\cite{rathee2021sirnn}. }. Therefore, we only present the principle and protocol of converting the shares from $\mathbb{Z}_{2^\ekk}$ to  $\mathbb{Z}_{2^\ell}$ in this paper.

To convert the shares of a secret $x$ from a small ring $\mathbb{Z}_{2^\ekk}$ to a large ring $\mathbb{Z}_{2^\ell}$, the key step is to compute the shares of $d_0$, $d_1$ and $\mathit{ovfl}$ on $\mathbb{Z}_{2^\ell}$ (i.e. $\share{d_0}_{2^\ell}$,$ \share{d_1}_{2^\ell}$, and $  \share{\mathit{ovfl}}_{2^\ell}$), where $d_0 = (\additiveshare{x}^0_{2^\ekk} + \additiveshare{x}^1_{2^\ekk}) \mod 2^\ekk$,  $d_1 = \additiveshare{x}^2_{2^\ekk}$, and $\mathit{ovfl} = d_0 + d_1 - x$.  After obtaining $\share{d_0}_{2^\ell}$,$ \share{d_1}_{2^\ell}$, and $  \share{\mathit{ovfl}}_{2^\ell}$, the parties can  compute the shares of $x$ on $\mathbb{Z}_{2^\ell}$ using $   \share{x}_{2^\ell} = \share{d_0}_{2^\ell} + \share{d_1}_{2^\ell} -  \share{\mathit{ovfl}}_{2^\ell}$. 


\begin{algorithm}
    \LinesNumbered
    \small
    \caption{$\mathit{ConvertShare}$}
    \label{pro:convertshare}
    \begin{flushleft}
 \textbf{Input:}    $\share{x}_{2^\ekk}$ $(x \in [0, 2^{\ekk-1})$, i.e.  $P_i$ inputs $\share{x}^i_{2^\ekk}=$ $(\additiveshare{x}^i_{2^\ekk}, \additiveshare{x}^{i+1}_{2^\ekk})$, and two public integers $\ekk$ and $\ell$ ($\ekk < \ell$).  \\
 \textbf{Output:}  $\share{x'}_{2^\ell}$, i.e. $P_i$ gets $\share{x'}^i_{2^\ell}=$ $(\additiveshare{x'}^i_{2^\ell}, \additiveshare{x'}^{i+1}_{2^\ell})$, with $x' = x$.  \\

 \textbf{Offline:}  $P_0, P_1, P_2$ generate dabit together, so $P_i$ holds $\additiveshare{r}^i_2$, $\additiveshare{r}^{i+1}_2$, $\additiveshare{r}^i_{2^\ell}$ and $\additiveshare{r}^{i+1}_{2^\ell}$ ( $r = 0 \ or \ 1$ ). \\
 \end{flushleft}

 \begin{algorithmic}[1]
 
     \STATE $P_0$ locally computes $d_0 =  (\additiveshare{x}^0_{2^\ekk} + \additiveshare{x}^1_{2^\ekk})\ mod \ 2^\ekk $.  $P_1$ locally computes $d_1 = \additiveshare{x}^2_{2^\ekk}$.
     \STATE $P_0$ locally computes $\mathit{truncd}_0 = d_0 \gg (\ekk-1)$.  $P_1$ locally computes  $ \mathit{truncd}_1 =  (-((-d_1) \gg {\ekk-1})) $.
     \STATE $P_0$ shares  $d_0$ and $\mathit{truncd}_0$ on $\mathbf{Z}_{2^\ell}$. $P_1$ shares $d_1$ and $\mathit{truncd}_1$ on $\mathbf{Z}_{2^\ell}$. 
     \STATE $\share{d}_{2^\ell} = \share{d_0}_{2^\ell} + \share{d_1}_{2^\ell}  $.
     \STATE $\share{\mathit{truncsum}}_{2^\ell} = \share{\mathit{truncd}_0}_{2^\ell} + \share{\mathit{truncd}_1}_{2^\ell}  $.

     \STATE  $P_0$ locally computes $b_0 = (\mathit{truncd}_0 \wedge 1)  \oplus  \additiveshare{r}^0_2 $ $ \oplus  \additiveshare{r}^1_2$.  $P_1$ locally computes $ b_1 = (\mathit{truncd}_1 \wedge 1) \oplus \additiveshare{r}^2_2$.
    \STATE $P_0$ sends $b_0$ to $P_1$ and $P_2$. $P_1$  sends $b_1$ to $P_0$ and $P_2$. 
    \STATE The parties all locally compute $b = b_0 \oplus b_1  $.
    \STATE $\share{bit}_{2^\ell} = b + \share{r}_{2^\ell} - 2 * b * \share{r}_{2^\ell}$.
    \STATE $\share{\mathit{ovfl}}_{2^\ell} = (\share{\mathit{truncsum}}_{2^\ell} - \share{bit}_{2^\ell}) * 2^{\ekk-1}  $.
    \STATE $\share{x'}_{2^\ell} = \share{d}_{2^\ell} - \share{\mathit{ovfl}}_{2^\ell} $.
 \end{algorithmic}
\end{algorithm}

$d_0$ can be computed and shared by $P_0$ ,since both $\additiveshare{x}^0_{2^\ekk}$ and $\additiveshare{x}^1_{2^\ekk}$ are held by $P_0$. $ d_1$ can be computed and shared by $P_1$, since $\additiveshare{x}^2_{2^\ekk}$ are held by $P_1$. Furthermore, the computation of $  \share{\mathit{ovfl}}_{2^\ell}$ is based on the following observation: all the secrets whose shares require conversion in the protocol \textit{ComputeModifiedGini} (Protocol~\ref{pro:computemodifiedgini}) are positive, i.e. the secrets belong to the range $[0, 2^{\ekk-1})$.  Additionally, $\mathit{ovfl} = 0 \ or \ 2^\ekk$, since $\mathit{ovfl} = (d_0 + d_1 - x) $ and $x = (d_0 + d_1) \ mod \ 2^\ekk $. Hence, to obtain $  \share{\mathit{ovfl}}_{2^\ell}$, the parties can first securely truncate the least significant $\ekk-1$ bits of $d_0 + d_1$, and then multiply the secret-shared truncated result with $2^{\ekk-1}$.

To securely truncate the least significant $\ekk-1$ bits of $d_0 + d_1$, we utilize Theorem~\ref{theorem1}, which is a variant of the Trunc Theorem proposed by Zhou et al.~\cite{zhou2022bicoptor}. According to Theorem~\ref{theorem1}, truncating $\ekk-1$ (i.e. $c = \ekk-1$) bits from $d_0$ and $d_1$ respectively may lead to one positive bit error in the truncated sum ($(d_0 \gg c) + (-((-d_1) \gg c))$) compared to the correct truncate result $\lfloor {d / 2^{\ekk-1}}  \rfloor$ ($d = d_0 + d_1)$. To eliminate this error, we observe that the least significant bit of $\lfloor {d / 2^{\ekk-1}} \rfloor$ is 0, since $d = x + \mathit{ovfl}$, $x \in [0, 2^{\ekk-1})$, and $\mathit{ovfl} = 0 \ or \ 2^\ekk$. This insight allows the parties to determine whether the truncated sum has one positive bit error based on its least significant bit. If the least significant bit is 1 after truncating $\ekk-1$ bits, one positive bit error occurs; if it is 0, the truncated sum is correct. Thus, the parties can compute the least significant bit of the truncated sum to eliminate its impact. 

In summary, to compute $\share{\mathit{ovfl}}_{2^\ell}$, $P_0$ and  $P_1$ first truncate $d_0$ and $d_1$ with $\ekk-1$ bits, respectively. This process removes the non-overflow portion and potentially introduces one positive bit error compared to the correct truncate result $\lfloor {d / 2^{\ekk-1}}  \rfloor$. Next, the parties securely compute the least significant bit of the truncated sum and eliminate the possible one positive bit error. Finally, the parties multiply the secret-shared truncated sum with $2^{\ekk-1}$, then get $  \share{\mathit{ovfl}}_{2^\ell}$.

\begin{theorem}
\label{theorem1}
 Let $c$ be an integer, satisfying $c < \ekk < \ell - 1 $. Let $d \in [0,  2^{\ekk+1})$, $d_0$ and $d_1$  $ \in [0, 2^\ekk)$ satisfying $d_0 + d_1 = d $. Then:
 
$ (d_0 \gg c) + (-((-d_1) \gg c)) = \lfloor {d / 2^c}  \rfloor + bit$,
where $bit =  0 \ or \ 1 $.

\begin{proof}
The proof is presented in Appendix~\ref{app.proof-of-corollary}.
\end{proof}

\end{theorem}

As is shown in Protocol~\ref{pro:convertshare}, the parties in the protocol \textit{ConvertShare} input a secret-shared value $\share{x}_{2^\ekk}$ ($x \in [0, 2^{\ekk-1})$)~\footnote{Note that the assumption $x \in [0, 2^{\ekk-1})$ is reasonable when we train decision trees since all the values need to be converted are all positive, although the studies~\cite{aly2019zaphod, damgaard2019new, gupta2022llama, jawalkar2023orca} adopt the assumption  $x \in [0, 2^{\ekk})$. } on the small ring $\mathbb{Z}_{2^\ekk}$ and two public integers $\ekk$ and $\ell$ ($\ekk < \ell$). The parties get a secret-shared value $\share{x'}_{2^\ell}$ on the large ring $\mathbb{Z}_{2^\ell}$, such that $x = x'$, as the output. 

During the offline phase of the protocol \textit{ConvertShare}, the parties generate a daBit~\cite{damgaard2019new}, which consists of the shares of a bit on both $\mathbb{Z}_2$ (a ring of size 2) and $\mathbb{Z}_{2^\ell}$. The daBit is used for masking the least significant bit of the truncated sum.

The online phase of the protocol \textit{ConvertShare} consists of three stages. (1) The first stage (Line 1-5): The parties compute  $\share{d}_{2^\ell}$ and the secret-shared truncated sum $\share{truncsum}_{2^\ell}$. To accomplish this, $P_0$ and $P_1$ first locally compute $d_0$ and $d_1$ (Line 1), and then truncate $d_0$ and $d_1$ to get $\mathit{truncd}_0$ and $\mathit{truncd}_1$ (Line 2). Next, $P_0$ and $P_1$ share these values on $\mathbb{Z}_{2^\ell}$ (Line 3). Finally, the parties compute $\share{d}_{2^\ell}$ and $\share{truncsum}_{2^\ell}$ by summing the secret-shared values (Line 4-5). (2) The second stage (Line 6-9): The parties compute the secret-shared one positive bit error $\share{bit}_{2^\ell}$.  To accomplish this, $P_0$ and $P_1$ first mask the least significant bits of $\mathit{truncd}_0$ and $\mathit{truncd}_1$ using the daBit (Line 6). Then, $P_0$ sends its masked bit to both $P_1$ and $P_2$. $P_1$ sends its masked bit to $P_0$ and $P_2$ (Line 7). Finally, the parties compute the secret-shared least significant bit $\share{bit}_{2^\ell}$ by XORing the two masked bits and then removing the mask (Line 8-9). (3) The third stage (Line 10-11): the parties compute the shares of the secret $x$ on $\mathbb{Z}_{2^\ell}$ by first computing the secret-shared overflow value $\share{ovfl}_{2^\ell}$, and then subtracting $\share{ovfl}_{2^\ell}$ from $\share{d}_{2^\ell}$. 


\subsection{Communication Complexity Comparison}

\label{sec.complexity}

We compare the communication complexity of \fname with two three-party frameworks~\cite{hamada2021efficient, MarkAbspoel2021SecureTO} for training a decision tree from two aspects: communication sizes and communication rounds. As shown in Table~\ref{tab.complexity}, \fname outperforms these two frameworks in both communication sizes and communication rounds. (1) Compared to the framework~\cite{MarkAbspoel2021SecureTO}, \fname reduces a exponential term $2^h$ to a linear term $h$ in the communication size complexity ($2^hmn \log n \ell \log \ell$ $\rightarrow$ $hmn \log n \ekk \log \ekk$ ). This improvement comes from the communication optimization based on the secure radix sort protocols~\cite{chida2019efficient} in \fname eliminates the need to train each node with a padded dataset. (2) Compared to the framework~\cite{hamada2021efficient}, \fname reduces a linear term $h$ in both communication sizes and communication rounds ($hmn\ell^2$ $\rightarrow$ $mn\ekk^2$ and $h \ell$ $\rightarrow$ $\ekk$). This improvement comes from that the communication optimization based on the secure radix sort protocols~\cite{chida2019efficient} eliminates the need to repeatedly generate permutations. (3) Compared to both frameworks~\cite{MarkAbspoel2021SecureTO, hamada2021efficient}, \fname reduces the dependency on the parameter $\ell$ to a smaller parameter $\ekk$ ($\ekk < \ell$) in both communication sizes and rounds. This improvement comes from the optimization based on our proposed share conversion protocol allows \fname to perform most computations on the small ring $\mathbb{Z}_{2^\ekk}$, rather than on the large ring $\mathbb{Z}_{2^\ell}$.


Besides, we present the communication complexity of the basic primitives and the detailed communication complexity analysis of each training protocol in \fname in Appendix~\ref{app.communication_complexity}.

\begin{table}[ht]
\centering
\caption{Communication complexity of \fname vs the frameworks~\cite{hamada2021efficient, MarkAbspoel2021SecureTO}. We assume $v < \log n \approx m < n$.   The communication complexity of these three frameworks is all analyzed based on the primitive's communication complexity presented in Appendix~\ref{app.communication_complexity}. }

\scalebox{0.8}{
\begin{tabular}{ccc}
\toprule
Framework                      & Sizes                                       & Rounds                          \\ \midrule
\fname                         & $O(hmn \log n \ekk \log \ekk + mn\ekk^2)$   & $O(h\log n \log \ekk + \ekk)$   \\ 
\cite{MarkAbspoel2021SecureTO} & $O(2^hmn \log n\ell \log \ell + mn\ell^2)$  & $O(h\log n \log \ell + \ell))$  \\ 
\cite{hamada2021efficient}     & $O(hmn \log n \ell \log \ell + hmn\ell^2 )$ & $O(h\log n \log \ell + h\ell))$ \\ 
\bottomrule
\end{tabular}
}
\label{tab.complexity}

\end{table}

\section{Performance Evaluation}
\label{sec.eva}


\subsection{Implementation and Experiment Settings}
\label{sec.experiment_setting}

\noindent\textbf{Implementation.}
We implement \fname based on \texttt{MP-SPDZ}~\cite{keller2020mp}, which is a widely used open-source framework for evaluating multi-party machine learning frameworks in the research field~\cite{lehmkuhl2021muse, dalskov2021fantastic, pentyala2021privacy, li2021privacy}. In \fname, computations are performed on a small ring $\mathbb{Z}_{ 2^\ekk }$ with $\ekk = 32$ and a large ring $\mathbb{Z}_{2^\ell}$ with $\ell = 128$. Besides, during the execution of the secure division operations, the bit length, $f$, for the decimal part is set to $2 \lceil \log n \rceil $, where $n$ is the number of samples in the training dataset.

\begin{table}[htbp]
\centering

\caption{Detailed information of datasets employed in our evaluation.}

\scalebox{0.8}{
\begin{tabular}{cccc}
\toprule
Dataset & \#Sample & \#Attribute & \#Label \\ 
\midrule
Kohkiloyeh   & 100           & 5                & 3            \\ 
Diagnosis    & 120           & 6                & 2            \\ 
Iris         & 150           & 4                & 3            \\ 
Wine         & 178           & 13               & 3            \\ 
Cancer       & 569           & 32               & 2            \\ 
Tic-tac-toe  & 958           & 9                & 2            \\ 
Adult        & 	
48,842         & 14               & 2            \\ 
Skin Segmentation        & 	
245,057         & 4               & 2            \\ 
\bottomrule
\end{tabular}}
\label{whole-datasets}

\end{table}

\noindent\textbf{Experiment Environment:}
We conduct experiments on a Linux server equipped with a 32-core 2.4 GHz Intel Xeon CPU and 512GB of RAM. Each party in \fname is simulated by a separate process with four threads. As for the network setting, we consider two scenarios: one is the LAN setting with a bandwidth of 1 gigabyte per second (GBps for short) and sub-millisecond round-trip time (RTT for short) latency. The other is the WAN setting with 5 megabytes per second (MBps for short)  bandwidth and 40ms RTT latency. We apply the tc tool\footnote{\url{https://man7.org/linux/man-pages/man8/tc.8.html}} to simulate these two network settings.

\noindent\textbf{Datasets:} As is shown in Table~\ref{whole-datasets}, we employ eight widely-used real-world datasets from the UCI repository~\cite{kelly2023uci} in our experiments. These datasets span a broad range in size, from the relatively small dataset, Kohkiloyeh, with only 100 samples, to the moderate-size dataset in the real world, Skin Segmentation, containing over 245,000 samples.

\subsection{Accuracy of \fname}

\begin{table}[htbp]
\centering
\caption{Accuracy of \fname vs. \texttt{scikit-learn} on eight widely used datasets. The height of decision trees is set to six.}

\scalebox{0.8}{
\begin{tabular}{ccc}
\toprule
Dataset      & \fname & \texttt{scikit-learn} \\ 
\midrule
Kohkiloyeh        & 0.7352 & 0.7352                \\ 
Diagnosis         & 1.0    & 1.0                   \\ 
Iris              & 0.9960 & 0.9607                \\ 
Wine              & 0.8622 & 0.8590                \\ 
Cancer            & 0.9388 & 0.9398                \\ 
Tic-tac-toe       & 0.8987 & 0.8987                \\ 
Adult             & 0.8494 & 0.8526                \\ 
Skin Segmentation & 0.9898 & 0.9898                \\ 
\bottomrule
\end{tabular}
}
\label{tab.accuracy}

\end{table}

We evaluate the accuracy of \fname by comparing it with the plaintext training algorithm for decision trees in \texttt{scikit-learn}~\cite{scikit-learn} on the datasets shown in Table~\ref{whole-datasets}. We divide each dataset into a training set and a test set with a ratio of $2:1$ and set the height of the decision trees to six. We conduct five runs for both \fname and \texttt{scikit-learn}. The average accuracy of the five runs is shown in Table~\ref{tab.accuracy}. As is shown in Table~\ref{tab.accuracy}, the accuracy of \fname and \texttt{scikit-learn} are almost the same. These accuracy results prove that \fname can accurately train a decision tree. Besides, we observe that the main reason for the small difference in the accuracy results is that when multiple split points have the same maximum modified Gini impurity, \fname and \texttt{scikit-learn} usually select different split points.

\subsection{Efficiency of \fname}

We compare the efficiency of \fname with two state-of-the-art three-party frameworks~\cite{MarkAbspoel2021SecureTO, hamada2021efficient}. These two frameworks are both based on RSS and adopt the same security model as \fname. Because these two frameworks are not open-sourced by the authors, for the framework proposed by Hamada et al.~\cite{hamada2021efficient}, we adopt the open-source implementation in \texttt{MP-SPDZ}, and modify the implementation to support multi-class datasets. For the framework proposed by Abspoel et al.~\cite{MarkAbspoel2021SecureTO}, we also implement it based on \texttt{MP-SPDZ}. Besides, we perform the computations of these two frameworks on the large ring $\mathbb{Z}_{2^\ell}$ ($\ell = 128$), so that these two frameworks could support datasets of almost the same size as \fname.

  The experimental results in Table~\ref{tab.efficiency} show that:

\begin{table*}[htbp]
\center
\caption{Online training time (seconds),  communication sizes (MBs), and communication rounds of \fname vs. two three-party  frameworks~\cite{MarkAbspoel2021SecureTO,hamada2021efficient} to train a decision tree with height six.  Note that the communication rounds, reported by the virtual machine of \texttt{MP-SPDZ}, are the cumulative totals across four threads. Therefore, the communication rounds exceed the actual number required.}

\scalebox{0.9}{
\begin{tabular}{cccccccccc}
\toprule
\multirow{2}{*}{}                                                                            & \multirow{2}{*}{Framework}     & \multicolumn{8}{c}{Dataset}                                                                                                                                                                                                                                                                                                                                                                                                                                                                                                                                                                                                                                                                                                                                                  \\ \cline{3-10} 
                                                                                             &                                & \multicolumn{1}{c}{Kohkiloyeh}                                                               & \multicolumn{1}{c}{Diagnosis}                                                                & \multicolumn{1}{c}{Iris}                                                                     & \multicolumn{1}{c}{Wine}                                                                     & \multicolumn{1}{c}{Cancer}                                                                    & \multicolumn{1}{c}{Tic-tac-toe}                                                              & \multicolumn{1}{c}{Adult}                                                                      & Skin Segmentation                                                          \\ \hline
\multirow{3}{*}{\begin{tabular}[c]{@{}c@{}}Online Training Time\\ in LAN\end{tabular}}       & \fname                         & \multicolumn{1}{c}{\textbf{\begin{tabular}[c]{@{}c@{}}1.5\\ ($5.2 \times$)\end{tabular}}}    & \multicolumn{1}{c}{\textbf{\begin{tabular}[c]{@{}c@{}}1.6\\ ($5.8 \times$)\end{tabular}}}    & \multicolumn{1}{c}{\textbf{\begin{tabular}[c]{@{}c@{}}1.1\\ ($5.3 \times$)\end{tabular}}}    & \multicolumn{1}{c}{\textbf{\begin{tabular}[c]{@{}c@{}}5.1\\ ($4.5 \times$)\end{tabular}}}    & \multicolumn{1}{c}{\textbf{\begin{tabular}[c]{@{}c@{}}23.3\\ ($5.2 \times$)\end{tabular}}}    & \multicolumn{1}{c}{\textbf{\begin{tabular}[c]{@{}c@{}}14.0\\ ($4.9 \times$)\end{tabular}}}   & \multicolumn{1}{c}{\textbf{\begin{tabular}[c]{@{}c@{}}1,187.8\\ ($4.1 \times$)\end{tabular}}}  & \textbf{\begin{tabular}[c]{@{}c@{}}1,783.5\\ ($3.5 \times$)\end{tabular}}  \\ \cline{2-10} 
                                                                                             & \cite{MarkAbspoel2021SecureTO} & \multicolumn{1}{c}{9.9}                                                                      & \multicolumn{1}{c}{11.4}                                                                     & \multicolumn{1}{c}{9.4}                                                                      & \multicolumn{1}{c}{33.1}                                                                     & \multicolumn{1}{c}{179.8}                                                                     & \multicolumn{1}{c}{108.5}                                                                    & \multicolumn{1}{c}{13,043.9}                                                                   & 18,601.9                                                                   \\ \cline{2-10} 
                                                                                             & \cite{hamada2021efficient}     & \multicolumn{1}{c}{7.8}                                                                      & \multicolumn{1}{c}{9.3}                                                                      & \multicolumn{1}{c}{5.9}                                                                      & \multicolumn{1}{c}{23.1}                                                                     & \multicolumn{1}{c}{122.3}                                                                     & \multicolumn{1}{c}{68.8}                                                                     & \multicolumn{1}{c}{4,933.2}                                                                    & 6,271.1                                                                    \\ \midrule
\multirow{3}{*}{\begin{tabular}[c]{@{}c@{}}Online Training Time\\ in WAN\end{tabular}}       & \fname                         & \multicolumn{1}{c}{\textbf{\begin{tabular}[c]{@{}c@{}}128.9\\ ($5.2 \times$)\end{tabular}}}  & \multicolumn{1}{c}{\textbf{\begin{tabular}[c]{@{}c@{}}129.3\\ ($5.2 \times$)\end{tabular}}}  & \multicolumn{1}{c}{\textbf{\begin{tabular}[c]{@{}c@{}}78.9\\ ($4.5 \times$)\end{tabular}}}   & \multicolumn{1}{c}{\textbf{\begin{tabular}[c]{@{}c@{}}302.3\\ ($4.5 \times$)\end{tabular}}}  & \multicolumn{1}{c}{\textbf{\begin{tabular}[c]{@{}c@{}}580.5\\ ($5.1 \times$)\end{tabular}}}   & \multicolumn{1}{c}{\textbf{\begin{tabular}[c]{@{}c@{}}238.6\\ ($5.0 \times$)\end{tabular}}}  & \multicolumn{1}{c}{\textbf{\begin{tabular}[c]{@{}c@{}}5,195.3\\ ($6.7 \times$)\end{tabular}}}  & \textbf{\begin{tabular}[c]{@{}c@{}}8,736.5\\ ($5.2 \times$)\end{tabular}}  \\ \cline{2-10} 
                                                                                             & \cite{MarkAbspoel2021SecureTO} & \multicolumn{1}{c}{679.8}                                                                    & \multicolumn{1}{c}{682.4}                                                                    & \multicolumn{1}{c}{420.2}                                                                    & \multicolumn{1}{c}{1,389.2}                                                                  & \multicolumn{1}{c}{3,397.0}                                                                   & \multicolumn{1}{c}{1,418.9}                                                                  & \multicolumn{1}{c}{70,168.4}                                                                   & 85,899.1                                                                   \\ \cline{2-10} 
                                                                                             & \cite{hamada2021efficient}     & \multicolumn{1}{c}{679.4}                                                                    & \multicolumn{1}{c}{681.3}                                                                    & \multicolumn{1}{c}{356.9}                                                                    & \multicolumn{1}{c}{1,374.8}                                                                  & \multicolumn{1}{c}{3,004.5}                                                                   & \multicolumn{1}{c}{1,210.3}                                                                  & \multicolumn{1}{c}{35,032.2}                                                                   & 42,315.1                                                                   \\ \hline
\multirow{3}{*}{\begin{tabular}[c]{@{}c@{}}Communication Sizes\\ (All parties)\end{tabular}} & \fname                         & \multicolumn{1}{c}{\textbf{\begin{tabular}[c]{@{}c@{}}24.9\\ ($9.3 \times$)\end{tabular}}}   & \multicolumn{1}{c}{\textbf{\begin{tabular}[c]{@{}c@{}}35.8\\ ($9.2 \times$)\end{tabular}}}   & \multicolumn{1}{c}{\textbf{\begin{tabular}[c]{@{}c@{}}34.1\\ ($9.2 \times$)\end{tabular}}}   & \multicolumn{1}{c}{\textbf{\begin{tabular}[c]{@{}c@{}}140.3\\ ($8.9 \times$)\end{tabular}}}  & \multicolumn{1}{c}{\textbf{\begin{tabular}[c]{@{}c@{}}980.7\\ ($9.2 \times$)\end{tabular}}}   & \multicolumn{1}{c}{\textbf{\begin{tabular}[c]{@{}c@{}}501.3\\ ($9.2 \times$)\end{tabular}}}  & \multicolumn{1}{c}{\textbf{\begin{tabular}[c]{@{}c@{}}51,725.3\\ ($8.7 \times$)\end{tabular}}} & \textbf{\begin{tabular}[c]{@{}c@{}}90,361.8\\ ($5.5 \times$)\end{tabular}} \\ \cline{2-10} 
                                                                                             & \cite{MarkAbspoel2021SecureTO} & \multicolumn{1}{c}{232.5}                                                                    & \multicolumn{1}{c}{331.9}                                                                    & \multicolumn{1}{c}{315.3}                                                                    & \multicolumn{1}{c}{1,251.8}                                                                  & \multicolumn{1}{c}{9,110.4}                                                                   & \multicolumn{1}{c}{4,637.0}                                                                  & \multicolumn{1}{c}{739,511.0}                                                                  & 875,991.0                                                                  \\ \cline{2-10} 
                                                                                             & \cite{hamada2021efficient}     & \multicolumn{1}{c}{276.3}                                                                    & \multicolumn{1}{c}{397.0}                                                                    & \multicolumn{1}{c}{336.3}                                                                    & \multicolumn{1}{c}{1,288.2}                                                                  & \multicolumn{1}{c}{9,734.3}                                                                   & \multicolumn{1}{c}{4,947.9}                                                                  & \multicolumn{1}{c}{450,378.0}                                                                  & 504,693.0                                                                  \\ \hline
\multirow{3}{*}{Communication Rounds}                                                        & \fname                         & \multicolumn{1}{c}{\textbf{\begin{tabular}[c]{@{}c@{}}17,526\\ ($4.6 \times$)\end{tabular}}} & \multicolumn{1}{c}{\textbf{\begin{tabular}[c]{@{}c@{}}20,882\\ ($4.5 \times$)\end{tabular}}} & \multicolumn{1}{c}{\textbf{\begin{tabular}[c]{@{}c@{}}15,931\\ ($4.3 \times$)\end{tabular}}} & \multicolumn{1}{c}{\textbf{\begin{tabular}[c]{@{}c@{}}54,472\\ ($3.9 \times$)\end{tabular}}} & \multicolumn{1}{c}{\textbf{\begin{tabular}[c]{@{}c@{}}111,242\\ ($5.1 \times$)\end{tabular}}} & \multicolumn{1}{c}{\textbf{\begin{tabular}[c]{@{}c@{}}33,914\\ ($5.1 \times$)\end{tabular}}} & \multicolumn{1}{c}{\textbf{\begin{tabular}[c]{@{}c@{}}61,638\\ ($5.3 \times$)\end{tabular}}}   & \textbf{\begin{tabular}[c]{@{}c@{}}15,142\\ ($4.8 \times$)\end{tabular}}   \\ \cline{2-10} 
                                                                                             & \cite{MarkAbspoel2021SecureTO} & \multicolumn{1}{c}{80,912}                                                                   & \multicolumn{1}{c}{96,042}                                                                   & \multicolumn{1}{c}{69,824}                                                                   & \multicolumn{1}{c}{216,326}                                                                  & \multicolumn{1}{c}{570,042}                                                                   & \multicolumn{1}{c}{174,948}                                                                  & \multicolumn{1}{c}{392,348}                                                                    & 98,030                                                                     \\ \cline{2-10} 
                                                                                             & \cite{hamada2021efficient}     & \multicolumn{1}{c}{106,286}                                                                  & \multicolumn{1}{c}{127,352}                                                                  & \multicolumn{1}{c}{86,300}                                                                   & \multicolumn{1}{c}{278,077}                                                                  & \multicolumn{1}{c}{655,562}                                                                   & \multicolumn{1}{c}{197,498}                                                                  & \multicolumn{1}{c}{331,412}                                                                    & 73,488                                                                     \\ \bottomrule
\end{tabular}
}
\label{tab.efficiency}

\end{table*}

\begin{itemize}[leftmargin=*]
  \item In the LAN setting,  \fname significantly outperforms the two frameworks by $3.5\times \sim 5.8\times$ in terms of training time.  The improvement primarily comes from that the two communication optimizations in \fname significantly reduce communication sizes, reaching an improvement of $5.5 \times \sim 9.3 \times$ compared to the baselines.

  \item  In the WAN setting, \fname significantly outperforms the baselines by $4.5\times \sim 6.7\times$ in terms of training time.  The improvement primarily comes from two facts. The first one is the same as that of the LAN setting, i.e. the two optimizations in \fname significantly reduce communication sizes. The second one is that the two optimizations in \fname also significantly reduce communication rounds, reaching an improvement of $3.9 \times \sim 5.3 \times$ compared to the baselines.

  \item Especially, \fname requires less than three hours (8,736.53s) in the WAN setting to train a decision tree on the dataset Skin Segmentation, which is a moderate-size dataset in the real world. This result shows that \fname is promising in the practical usage of privacy preserving training for decision trees. 

\end{itemize}

\subsection{Effectiveness of Communication Optimizations}

In order to further show the effectiveness of the two communication optimizations in \fname, we implement the two communication optimizations in Hamada et al.'s framework~\cite{hamada2021efficient}. We name Hamada et al.'s framework with the communication optimization based on the secure radix sort protocols~\cite{chida2019efficient} as Hamada et al.'s framework-radixsort, and Hamada et al.'s framework with the communication optimization based on our proposed share conversion protocol as Hamada et al.'s framework-convert.  We compare the training time, communication sizes, and communication rounds of \fname, these two frameworks, and the original Hamada et al.'s framework.

\begin{figure}[htbp]
\centering  
\subfigbottomskip=-5pt 
\subfigcapskip=-5pt 
\subfigure[Training Time in LAN]{
\includegraphics[width=0.23\textwidth]{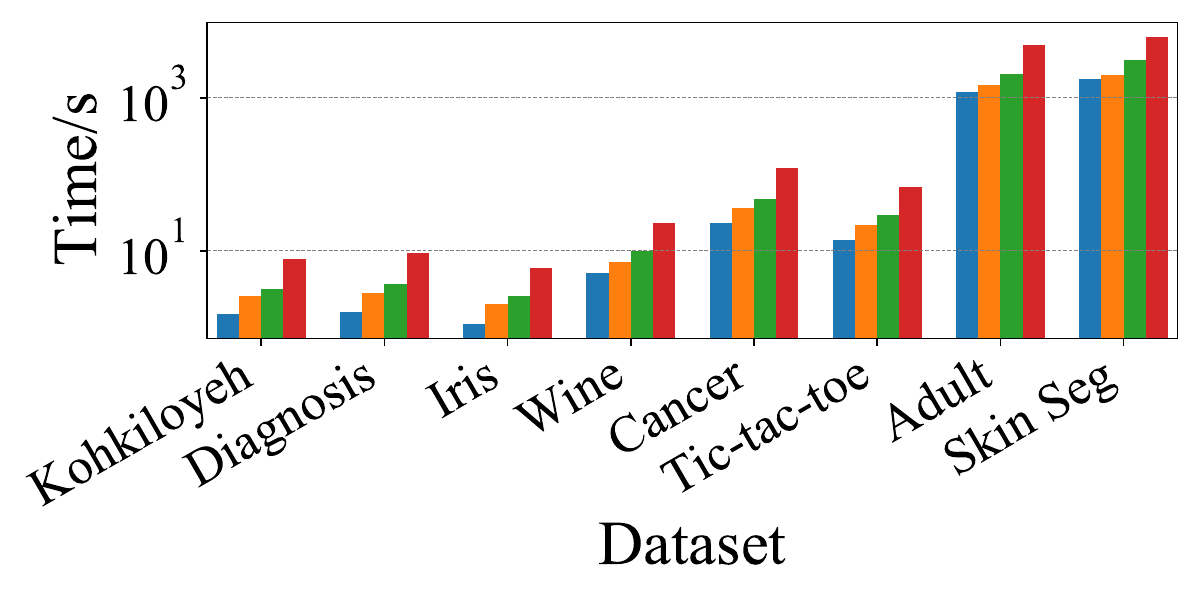}}
\hspace{-2mm}
\subfigure[Training Time in WAN]{
\includegraphics[width=0.23\textwidth]{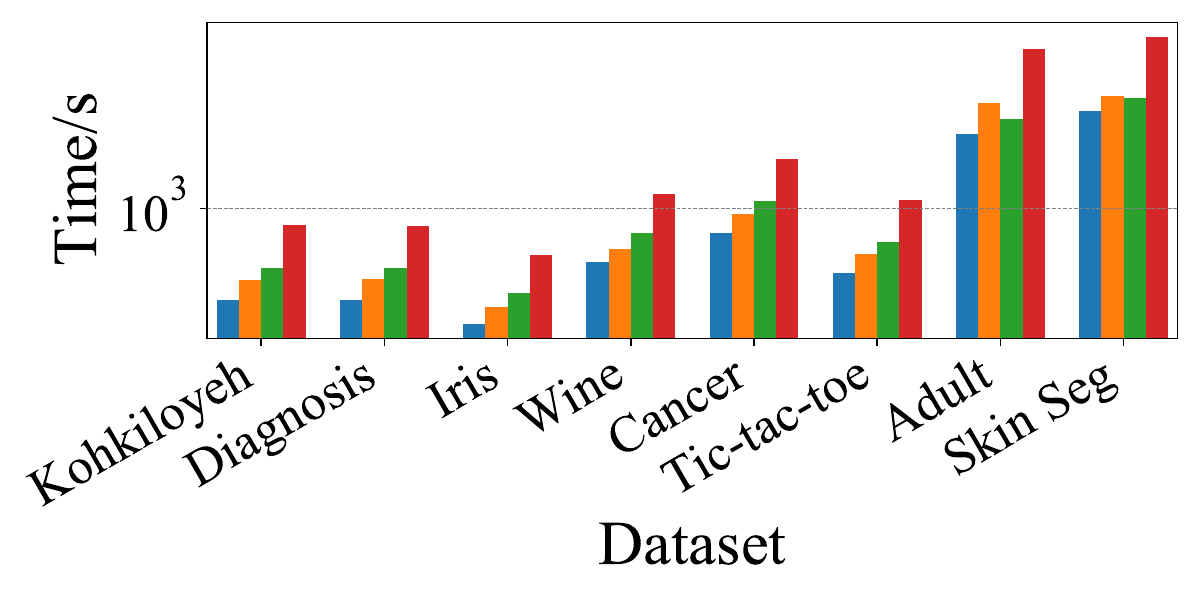}}

\subfigure[Communication Sizes]{
\includegraphics[width=0.23\textwidth]{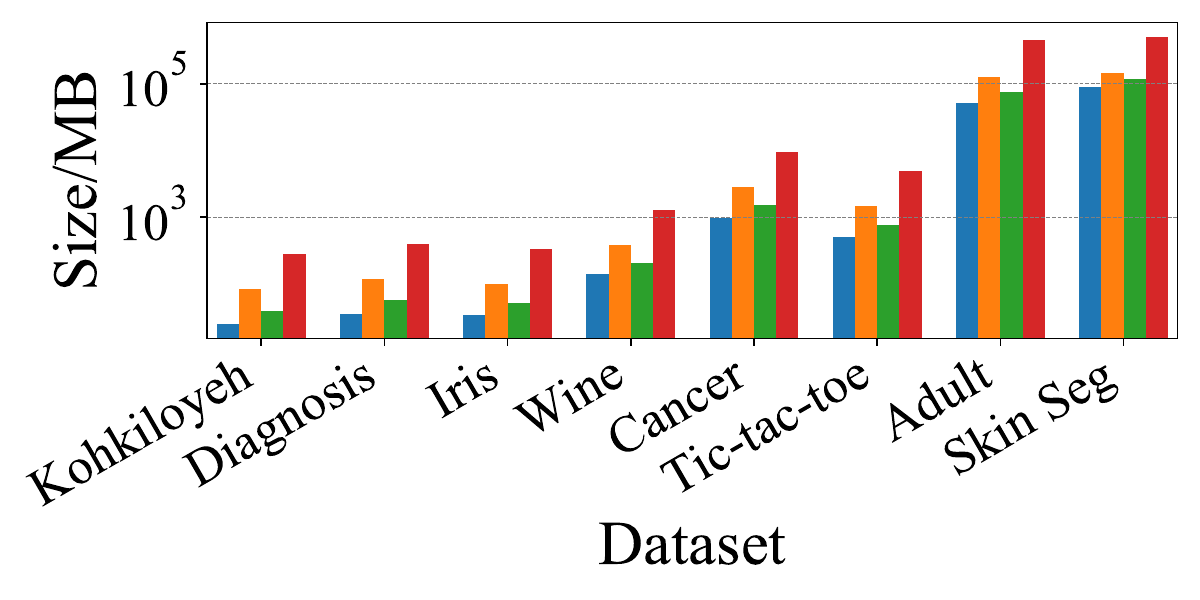}}
\hspace{-2mm}
\subfigure[Communication Rounds]{
\includegraphics[width=0.23\textwidth]{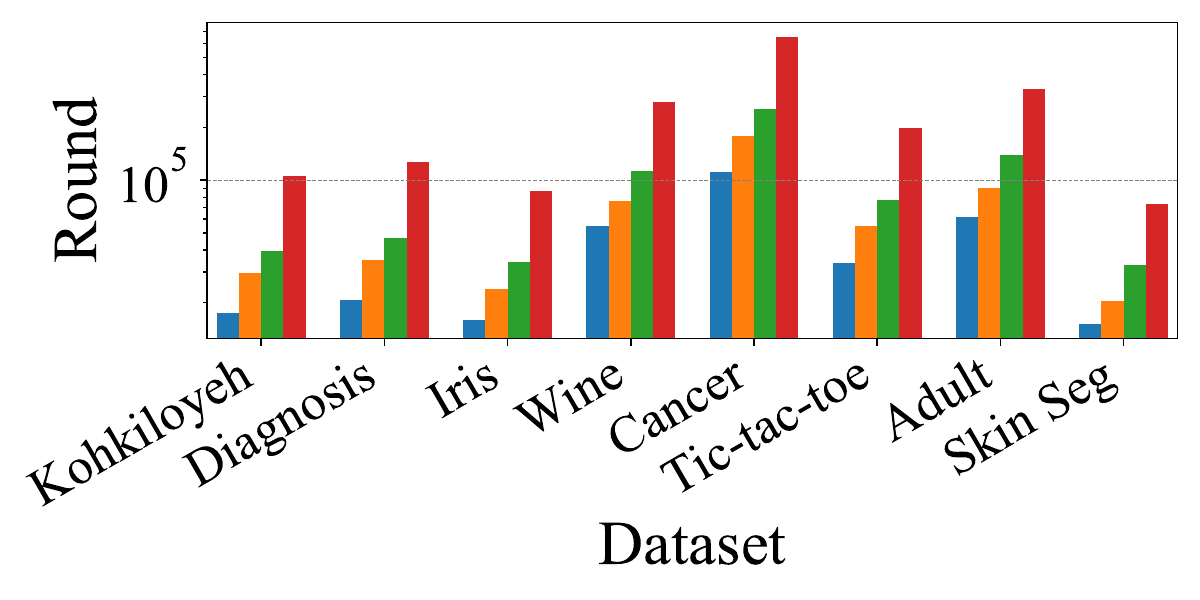}}

\subfigure{
\includegraphics[width=0.45\textwidth]{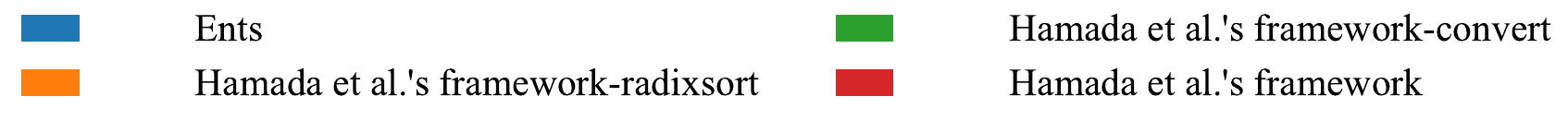}}

\caption{Online training time (seconds), communication sizes (MBs), and communication rounds of \fname, Hamada et al.'s framework-radixsort, Hamada et al.'s framework-convert, and Hamada et al.'s framework. `Skin Seg' refers to Skin Segmentation.}

\label{fig.op-effciency}
\end{figure}

As is shown in Figure~\ref{fig.op-effciency}, both the communication optimizations significantly improve the efficiency of Hamada et al.'s framework. (1) The communication optimization based on the secure radix sort protocols~\cite{chida2019efficient} improves the training time of Hamada et al.'s framework by $2.9 \times \sim 3.4 \times$ in the LAN setting and $3.1 \times \sim 3.7 \times$ in the WAN setting. In terms of communication sizes and communication rounds, this communication optimization yields improvements of $3.2 \times \sim 3.5 \times$ and $3.5 \times \sim 3.6 \times$, respectively. The improvements are primarily because this communication optimization enables Hamada et al.'s framework-radixsort to securely generate permutations only once, while the original Hamada et al.'s framework requires securely generating permutations when training each internal layer of a decision tree. (2) The communication optimization based on our proposed share conversion protocol improves the training time of Hamada et al.'s framework by $1.9 \times \sim 2.5 \times$ in the LAN setting and $2.3 \times \sim 4.7 \times$ in the WAN setting. In terms of communication sizes and communication rounds, this communication optimization yields improvements of $4.2 \times \sim 7.0 \times$ and $2.2 \times \sim 2.7 \times$, respectively. The improvements are primarily because this communication optimization enables Hamada et al.'s framework-convert to perform most computations on the small ring to reduce huge communication size. Additionally, this optimization also reduces communication sizes and communication rounds for the secure generation of permutations, because the secure generation of permutations can also be performed on the small ring. 

With the two communication optimizations, \fname significantly outperforms Hamada et al.'s framework in terms of the training time by $3.5 \times \sim 5.8 \times$ in the LAN setting, $4.5 \times \sim 6.7 \times$ in the WAN setting. In terms of communication sizes and communication rounds, \fname significantly outperforms Hamada et al.'s framework by $5.5 \times \sim 11.0 \times$ and $4.8 \times \sim 6.0 \times$, respectively.

Note that although Hamada et al.’s framework-radixsort requires more communication sizes than Hamada et al.’s framework-convert, Hamada et al.’s framework-radixsort still outperforms Hamada et al.’s framework-convert in the LAN setting. The primary reason for this phenomenon is that the communication optimization based on the secure radix sort protocols~\cite{chida2019efficient} also saves a lot of memory access time for Hamada et al.’s framework-radixsort. The process of generating permutations requires a lot of random memory access, which is very time-consuming. The communication optimization based on the secure radix sort protocols~\cite{chida2019efficient} enables Hamada et al.’s framework-radixsort to generate permutations only once. Thus, the memory access time of Hamada et al.’s framework-radixsort is much less than Hamada et al.’s framework-convert.


\subsection{Efficiency of Conversion Protocol}

We compare the efficiency of our proposed share conversion protocol, \textit{ConvertShare} (Protocol~\ref{pro:convertshare}), against three state-of-the-art conversion protocols: (1) the conversion protocol based on arithmetic and boolean share conversion (\textit{Convert-A2B} for short)~\cite{mohassel2018aby3}. (2) the conversion protocol based on Dabits (\textit{Convert-Dabits} for short)~\cite{aly2019zaphod}. (3) the conversion protocol based on function secret sharing (\textit{Convert-FSS} for short)~\cite{gupta2022llama}. We implement all these protocols in \texttt{MP-SPDZ}~\cite{keller2020mp}


To convert the replicated shares of a secret from $\mathbb{Z}_{2^{32}}$ to $\mathbb{Z}_{2^{128}}$, the communication sizes and communication rounds of the above conversion protocols are as follows: (1) our proposed protocol \textit{ConvertShare} (Protocol~\ref{pro:convertshare}) requires a communication size of $516$ bits in one communication round. (2) the protocol \textit{Convert-A2B}  requires a communication size of $1,506$ bits in six communication rounds. (3) the protocol \textit{Convert-Dabits}  requires a communication size of $22,860$ bits in three communication rounds. (4) the protocol \textit{Convert-FSS} requires a communication size of $512$ bits in two communication rounds. Besides, to convert the replicated shares of a vector of size $l$ from $\mathbb{Z}_{2^{32}}$ to $\mathbb{Z}_{2^{128}}$, the communication sizes of the above protocols will scale up by a factor of $l$. While the number of communication rounds remains constant.

 We evaluate the online runtime of the above conversion protocols for converting secret-shared vectors of size 1, 10, 100, 1000, and 10000 in both LAN and WAN settings. As is shown in Table~\ref{tab.convert_runtime}, our proposed protocol \textit{ConvertShare} significantly outperforms the other protocols by $2.0\times \sim 4.8\times$ in the LAN setting and by $2.0\times \sim 4.7\times$ in the WAN setting. This advancement is primarily due to its lower communication sizes and fewer communication rounds.   Note that although the protocol \textit{Convert-FSS} has a marginally smaller communication size than our proposed protocol \textit{ConvertShare}, the protocol \textit{Convert-FSS} burdens significantly higher local computation overhead, owing to its reliance on function secret sharing. Therefore, our proposed protocol \textit{ConvertShare} is more efficient than the protocol \textit{Convert-FSS} in both the LAN and WAN settings.


\begin{table}[htbp]
\centering
\caption{Online runtime (milliseconds) of conversion protocols for converting secret-shared vectors of size 1, 10, 100, 1000, and 10000 from $\mathbf{Z}_{2^{32}}$ to $\mathbf{Z}_{2^{128}}$.}

\scalebox{0.8}{
\begin{tabular}{ccccccc}
\toprule
\multirow{2}{*}{}    & \multirow{2}{*}{Protocol}                    & \multicolumn{5}{c}{Vector Size}                                                                                                                                                                                                                                                                                                                                                                                                                                \\ \cline{3-7} 
                     &                                              & \multicolumn{1}{c}{1}                                                                      & \multicolumn{1}{c}{10}                                                                     & \multicolumn{1}{c}{100}                                                                    & \multicolumn{1}{c}{1,000}                                                                  & 10,000                                                                  \\ \midrule
\multirow{4}{*}{LAN} & \textit{ConvertShare}                        & \multicolumn{1}{c}{\textbf{\begin{tabular}[c]{@{}c@{}}0.58\\ ($ 2.0\times$)\end{tabular}}} & \multicolumn{1}{c}{\textbf{\begin{tabular}[c]{@{}c@{}}0.59\\ ($ 4.5\times$)\end{tabular}}} & \multicolumn{1}{c}{\textbf{\begin{tabular}[c]{@{}c@{}}1.2\\ ($ 4.8\times$)\end{tabular}}}  & \multicolumn{1}{c}{\textbf{\begin{tabular}[c]{@{}c@{}}6.2\\ ($ 4.1\times$)\end{tabular}}}  & \textbf{\begin{tabular}[c]{@{}c@{}}57.0\\ ($ 3.3\times$)\end{tabular}}  \\ \cline{2-7} 
                     & \textit{Convert-A2B}~\cite{mohassel2018aby3} & \multicolumn{1}{c}{2.6}                                                                    & \multicolumn{1}{c}{2.7}                                                                    & \multicolumn{1}{c}{5.8}                                                                    & \multicolumn{1}{c}{26.7}                                                                   & 191.1                                                                   \\ \cline{2-7} 
                     & \textit{Convert-Dabits}~\cite{aly2019zaphod} & \multicolumn{1}{c}{1.2}                                                                    & \multicolumn{1}{c}{4.5}                                                                    & \multicolumn{1}{c}{16.2}                                                                   & \multicolumn{1}{c}{137.4}                                                                  & 1,310.8                                                                 \\ \cline{2-7} 
                     & \textit{Convert-FSS}~\cite{gupta2022llama}   & \multicolumn{1}{c}{5.0}                                                                    & \multicolumn{1}{c}{19.1}                                                                   & \multicolumn{1}{c}{113.0}                                                                  & \multicolumn{1}{c}{1,103.9}                                                                & 11,004.5                                                                \\ \hline
\multirow{4}{*}{WAN} & \textit{ConvertShare}                        & \multicolumn{1}{c}{\textbf{\begin{tabular}[c]{@{}c@{}}40.6\\ ($ 2.0\times$)\end{tabular}}} & \multicolumn{1}{c}{\textbf{\begin{tabular}[c]{@{}c@{}}40.8\\ ($ 2.0\times$)\end{tabular}}} & \multicolumn{1}{c}{\textbf{\begin{tabular}[c]{@{}c@{}}42.0\\ ($ 4.1\times$)\end{tabular}}} & \multicolumn{1}{c}{\textbf{\begin{tabular}[c]{@{}c@{}}60.4\\ ($ 4.7\times$)\end{tabular}}} & \textbf{\begin{tabular}[c]{@{}c@{}}202.4\\ ($ 2.7\times$)\end{tabular}} \\ \cline{2-7} 
                     & \textit{Convert-A2B}~\cite{mohassel2018aby3} & \multicolumn{1}{c}{243.3}                                                                  & \multicolumn{1}{c}{243.7}                                                                  & \multicolumn{1}{c}{247.6}                                                                  & \multicolumn{1}{c}{284.5}                                                                  & 563.5                                                                   \\ \cline{2-7} 
                     & \textit{Convert-Dabits}~\cite{aly2019zaphod} & \multicolumn{1}{c}{141.7}                                                                  & \multicolumn{1}{c}{144.7}                                                                  & \multicolumn{1}{c}{173.1}                                                                  & \multicolumn{1}{c}{519.3}                                                                  & 3,394.3                                                                 \\ \cline{2-7} 
                     & \textit{Convert-FSS}~\cite{gupta2022llama}   & \multicolumn{1}{c}{84.7}                                                                   & \multicolumn{1}{c}{84.9}                                                                   & \multicolumn{1}{c}{180.9}                                                                  & \multicolumn{1}{c}{1,711.9}                                                                & 10,979.5                                                                \\ \bottomrule
\end{tabular}
}

\label{tab.convert_runtime}

\end{table}
\label{evaluation}

\section{Discussion}
\label{sec.diss}


\noindent\textbf{Discrete Attributes in Training Datasets.}
Although the protocols (Protocol~\ref{pro:traindecisiontree},~\ref{pro:traininternallayer},~\ref{pro:computemodifiedgini},~\ref{pro:attributewisesplitselection},~\ref{pro:testsamples},~\ref{pro:trainleaflayer}) in our proposed \fname only process continuous attributes in training datasets,  these protocols can be modified to process discrete attributes. According to the technical routine introduced by Abspoel et al.~\cite{MarkAbspoel2021SecureTO}, the main modification to the above training protocols to process discrete attributes are as follows: (1) The procedure of generating and applying permutations should be removed, because training on discrete attributes does not require sorting the samples according to the attributes. (2) The less-than tests between the samples and split threshold should be replaced with equality tests, because the splitting test for discrete attributes requires checking whether the attribute of samples is equal to the split threshold, rather than checking whether the attribute of samples is less than the split threshold.

\noindent\textbf{\fname in Two-Party Scenarios.} \fname can be adapted to train a decision tree in two-party scenarios by modifying the basic MPC protocols introduced in Section~\ref{sec.mpc} accordingly. Note that except for our proposed conversion protocol \textit{ConvertShare},  all the other basic protocols are already supported for the two-party semi-honest security model in \texttt{MP-SPDZ}.  Thus, we provide the modification of our proposed protocol \textit{ConvertShare} to two-party scenarios in Appendix~\ref{appendix.twopartyextend}. 

To evaluate the efficiency of two-party \fname,  we perform performance evaluation for two-party \fname and two state-of-the-art two-party frameworks~\cite{liu2020towards, akavia2022privacy}. The experimental results, provided in Appendix~\ref{appendix.twopartyexperiment},  show that two-party \fname also significantly outperforms the two-party frameworks~\cite{liu2020towards, akavia2022privacy}, with a training time improvement of $14.3\times \sim 1,362\times$ in the LAN setting and $4.7\times \sim 6.4\times$ in the WAN setting.

\noindent\textbf{Security of \fname.} \fname is designed to be secure under a three-party semi-honest security model with an honest majority. Note that except for the protocol \textit{ConvertShare}, other protocols in \fname are constructed using the security-proven protocols introduced in Section~\ref{sec.mpc}.  To prove the security of the protocol \textit{ConvertShare},  we provide a security analysis using the standard real/ideal world paradigm in Appendix~\ref{app.securityanalysis}. 


\noindent\textbf{Practical Usages of \fname.}
\fname is promising in the practical usage of privacy preserving training for decision trees. According to the experimental results, \fname costs less than three hours to train a decision tree on a real-world dataset with more than 245,000 samples in the WAN setting. Besides, with a larger bandwidth and more powerful servers, the training time of \fname can be further reduced.

\noindent\textbf{Future Work.} In the future, we will support more tree-based models, such as XGBoost~\cite{chen2016xgboost}, in \fname. Because decision trees are the basic components of these tree-based models, the optimizations proposed in this paper should also be suitable for them.

\section{Related Work}
\label{sec.rel}

\noindent \textbf{Multi-party training frameworks for decision trees.}
Over the last decade, many multi-party training frameworks~\cite{dansana2013novel,  li2019outsourced, sheela2013novel, liu2020towards, lu2023squirrel, MarkAbspoel2021SecureTO, JaideepVaidya2014ARD, hamada2021efficient, akavia2022privacy, adams2022privacy} for decision trees have been proposed. These frameworks can be categorized into two types according to whether they disclose some private information during the training process. The first type frameworks~\cite{ dansana2013novel,  li2019outsourced, sheela2013novel, liu2020towards, lu2023squirrel} disclose some private information, such as the split points of decision tree nodes and Gini impurity, and rely on the disclosed private information to speed up the training process. For instance, the framework proposed by Lu et al.~\cite{lu2023squirrel} discloses the split points of decision tree nodes to a specific party, so that some comparisons between the samples and the split thresholds can be performed locally to speed up the training process. However, Zhu et al.~\cite{zhu2010understanding} demonstrate that the disclosed information may be exploited to infer training data. Thus, the frameworks of this type may be unsuitable in scenarios where training data must be strictly protected.

The second type frameworks~\cite{MarkAbspoel2021SecureTO, JaideepVaidya2014ARD, hamada2021efficient, akavia2022privacy, adams2022privacy}, on the other hand, does not disclose private information during the training process but suffer from low accuracy or inefficiency. Vaidya et al.\cite{JaideepVaidya2014ARD} and Adams et al.\cite{adams2022privacy} both propose frameworks for securely training random decision trees. However, random decision trees usually have lower accuracy compared to conventional decision trees. Abspoel et al.\cite{MarkAbspoel2021SecureTO} and Hamada et al.\cite{hamada2021efficient} both propose three-party training frameworks based on RSS for decision trees. However, these frameworks are inefficient due to high communication overhead. Akavia et al.~\cite{akavia2022privacy} introduced a two-party framework based on homomorphic encryption. However, this framework is also inefficient due to high computation overhead. Due to limitations in accuracy or efficiency, these frameworks of this type remain impractical.

\noindent \textbf{Shares conversion protocols.}
Recently, several share conversion protocols have been proposed~\cite{kelkar2022secure, mohassel2018aby3, aly2019zaphod, damgaard2019new, gupta2022llama, jawalkar2023orca}. Kelkar et al.~\cite{kelkar2022secure} introduce a share conversion protocol that does not require interaction but suffers from a high failure probability, making it unsuitable for training decision trees. The arithmetic and boolean conversion protocol proposed by Mohassel and Rindal~\cite{mohassel2018aby3} can be used to implement the correct share conversion between a small ring and a large ring. However,  this method is inefficient since it requires six communication rounds.  Aly et al.\cite{aly2019zaphod}  propose a share conversion protocol based on Dabits~\cite{damgaard2019new}, but this protocol requires considerable communication sizes and three communication rounds, which make it inefficient too.  Gupta et al.\cite{gupta2022llama} and Jawalkar et al.~\cite{jawalkar2023orca} both propose share conversion protocols based on function secret sharing~\cite{boyle2016function}, requiring zero and one communication rounds, respectively. However, when combining function secret sharing with replicated secret sharing, two additional communication rounds are introduced: one for masking shares and another for resharing shares. Therefore, Gupta's method requires two communication rounds, and Jawalkar's method requires three communication rounds. Besides, these two methods also suffer from high computation overhead due to dependence on function secret sharing.  In summary, the above protocols either suffer from high failure probability or lack efficiency.

\noindent \textbf{Our advantages.} (1) Our proposed framework \fname does not disclose private information during the training process, and meanwhile, it attains the same accuracy level as the plaintext training algorithm for decision trees in \texttt{scikit-learn} and significantly outperforms the existing frameworks that also do not disclose private information during the training process in terms of efficiency. (2) Our proposed conversion protocol demonstrates superior efficiency while guaranteeing its correctness.

\section{Conclusion}
\label{sec.con}

In this paper, we propose \fname, an efficient three-party training framework for decision trees, and optimize the communication overhead from two-folds: (1) We present a series of training protocols based on the secure radix sort protocols~\cite{chida2019efficient} to efficiently and securely split a dataset with continuous attributes. (2) We propose an efficient share conversion protocol to convert shares between a small ring and a large ring to reduce the huge communication overhead incurred by performing almost all computations on a large ring.  Experimental results from eight widely used datasets demonstrate that \fname outperforms state-of-the-art frameworks by $5.5\times \sim 9.3\times$ in communication sizes and $3.9\times \sim 5.3\times$ in communication rounds. In terms of training time,
\fname yields an improvement of $3.5\times \sim 6.7\times$. Especially, \fname requires less than three hours to train a decision tree on a  real-world dataset (Skin Segmentation) with more than 245,000 samples in the WAN setting. These results show that \fname is promising in the practical usage of privacy preserving training for decision trees.

\section*{Acknowledgement}
This paper is supported by the National Key R\&D Program of China (2023YFC3304400), Natural Science Foundation of China (62172100, 92370120). We thank all anonymous reviewers for their insightful comments. Weili Han is the corresponding author.

\bibliographystyle{ACM-Reference-Format}

\bibliography{reference}

\appendix

\section{Notations Used in This Paper}
\label{app.notations}

 As is shown in Table~\ref{tab:notations}, we list the notations used in this paper for clarity purposes.

\begin{table}[ht]
\centering
\caption{Notations used in this paper.}

\scalebox{0.85}{
\begin{tabular}{cc} 
\toprule
Notation                                      & Description                                                                                                                                                                                                     \\ 
\midrule
$\mathcal{D}$                                      & The dataset for training a decision tree.                                                                                                                                                                                                     \\ 
\hline
$\mathcal{D}_{condition}$                            & \begin{tabular}[c]{@{}c@{}} A subset of $\mathcal{D}$. It contains \\ all the samples satisfying the given condition in $\mathcal{D}$. \end{tabular}                                                                                            \\ 

\hline
$\mathcal{D}^{node}$                            & \begin{tabular}[c]{@{}c@{}} The dataset for training a node.  \\ It is a subset of  $\mathcal{D}$. \end{tabular}                                                                                            \\ 
\hline
$n$                                                & The number of samples in the training dataset.                                                                                                                                                                              \\ 
\hline
$m$                                                & The number of attributes of a sample.                                                                                                                                                                           \\ 
\hline
$v$                                                & \begin{tabular}[c]{@{}c@{}}The number of labels.\end{tabular}                                                                                                        \\ 
\hline
$t$                                                & The split threshold of a node.                                                                                                                                                                                                 \\ 
\hline
$a_0,\cdots,a_{m-1}$                                    & The $m$ attributes of a sample.                                                                                                                                                                                           \\ 
\hline
$y$                                                & The label of a sample.                                                                                                                                                                                                  \\ 
\hline

$h$                                                & The height of a decision tree.                                                                                                                                                                                            \\ 
\hline
$P_i$                                              & \begin{tabular}[c]{@{}c@{}}Party $i$~participating in the secure \\decision tree training, where $i \in \{0,1,2\}$.\end{tabular}                                                                                        \\ 
\hline

$\mathbb{Z}_2$ , $\mathbb{Z}_{ 2^\ekk } ,\mathbb{Z}_{2^\ell}$                                   & A ring of size $2,  2^\ekk  ,2^\ell  (2 < \ekk < \ell )$ respectively.                                                                                                                                                                                                                                                                                                                                                    \\ 
\hline
$f$                                              & \begin{tabular}[c]{@{}c@{}}The bit length for \\ representing the decimal part.   \end{tabular}                                                                                                                      
                               \\ 
\hline
$\share{x}$                                      & \begin{tabular}[c]{@{}c@{}}The replicated shares of a secret $x$, \\ i.e. $P_i$ holds $\share{x}^i=$ $( \additiveshare{x}^i, \additiveshare{x}^{i+1})$    \end{tabular}   

\\
\hline

$\vec{a_i}$                                        &
\begin{tabular}[c]{@{}c@{}} The~$i$-th attribute vector that contains \\ the $i$-th attribute values of all the samples, i.e. $\vec{a_i}[j]$  \\ represents the value of $i$-th attribute of~$j$-th sample.~\end{tabular}                                                                                   \\ 
\hline
$\vec{y}$                                        & \begin{tabular}[c]{@{}c@{}} The label vector that contains \\ the label of all the samples, i.e. $\vec{y}[i]$  \\ represents the label of of~$i$-th sample.\end{tabular}                                                                                   \\
\hline

$\vec{\pi},\vec{\alpha}$ & Permutations used to reorder vectors.                                                                                                                                                                                  \\ 
\hline
$\vec{g}$                                          &           \begin{tabular}[c]{@{}c@{}} The group flag vector \\ used to indicates group boundaries.\end{tabular}                                                                                                                                             \\
\hline

$\vv{\mathit{spnd}}$                                        & \begin{tabular}[c]{@{}c@{}}A sample-node vector.  \\ The equation $\vv{\mathit{spnd}}[j]=i$  represents that \\ the $j$-th sample  belongs to the node whose $nid$ is $i$. \end{tabular}                                                                                  \\
\hline

$\vv{\mathit{spat}}$                                        & \begin{tabular}[c]{@{}c@{}} A sample-attribute vector.  \\  The equation $\vv{\mathit{spat}}[j]=i$ represents that \\ the split attribute index of the node \\ to which the $j$-th sample  belongs  is $i$.\end{tabular}                                                                                 \\
\hline
$\vv{\mathit{spth}}$                                        & \begin{tabular}[c]{@{}c@{}}A sample-threshold vector . \\ The equation $\vv{\mathit{spth}}[j]=i$  represents  that \\ the split threshold of the node \\ to which  the $j$-th sample  belongs   is $i$.\end{tabular}                                                                               \\
\hline

$\vv{\mathit{spnd}}^{(k)}$                                        & The sample-node vector in the $k$-th layer.                                                                          \\
\hline

$\vv{\mathit{spat}}^{(k)}$                                        &  The sample-attribute vector in the $k$-th layer.                                                                                  \\
\hline
$\vv{\mathit{spth}}^{(k)}$                                        & The sample-threshold vector in the $k$-th layer.                                                                              \\
\hline
$\vv{\mathit{spth}}_i$                                        &  \begin{tabular}[c]{@{}c@{}}The sample-threshold vector \\ computed with only the $i$-th attribute. \end{tabular}                                                                               \\
\hline

$\mathit{MinValue}$ & The minimum value on the ring $\mathbb{Z}_{2^\ekk}$                                                                                                                                                                                  \\ 

\bottomrule

\end{tabular}

}
\label{tab:notations}

\end{table}


\section{Remaining Protocols For Training Process}

\label{appendix.pro}







The protocol \textit{AttributeWiseSplitSelection} used to compute the modified Gini impurity and split threshold with an attribute $a_i$ is shown in Protocol~\ref{pro:attributewisesplitselection}. The parties in this protocol input a secret-shared group flag vector $\share{\vec{g}}_{ 2^\ekk }$, a secret-shared attribute vector $\share{\vec{a}_i}_{ 2^\ekk }$, and a secret-shared label vector $\share{\vec{y}}_{ 2^\ekk }$. They get a secret-shared sample-threshold vector $\share{\vv{\mathit{spth}}_i}_{ 2^\ekk }$, which are computed with only the attribute $a_i$, and a secret-shared modified Gini impurity vector $\share{\vv{\mathit{gini}}_i}_{ 2^\ekk }$  as the output. The attribute vector and the label vector are assumed to have been sorted according to $\vec{a_i}$ within each node. 

The protocol \textit{AttributeWiseSplitSelection} consists of three stages: (1) The first stage (Line 1-3):  the parties compute the modified Gini impurity and the split threshold for each split point $(a_i, \vec{\mathit{t}}[j])$, where $\vec{\mathit{t}}[j] = (\vec{a}_i[j] + \vec{a}_i[j+1])/2$ for each $j \in [0, n-2]$ and $\vec{\mathit{t}}[n-1] = \vec{a}_i[n-1]$. (2) The second stage (Line 4-6):  the parties check whether the $i$-th sample is the final sample of a node. If it is true, the modified Gini impurity is set to the $\mathit{MinValue}$ ($\mathit{MinValue}$ refers to the minimum value in the ring $\mathbb{Z}_{2^\ekk}$) to avoid splitting between the $i$-th sample and $(i+1)$-th sample. (3) The third stage (Line 7): the parties call the protocol \textit{GroupMax} (introduced in Section~\ref{sec.group}) to obtain the secret-shared sample-threshold for each node and the secret-shared maximum modified Gini impurity. 

Note that this protocol is the same as the protocol proposed by Hamada et al.~\cite{hamada2021efficient}.

\begin{algorithm}
    \LinesNumbered
    \small
    \caption{$\mathit{AttributeWiseSplitSelection}$}
    \label{pro:attributewisesplitselection}
    \begin{flushleft}
 \textbf{Input:} A secret-shared group flag  vector $\share{\vec{g}}_{ 2^\ekk }$, a secret-shared attribute vector $\share{\vec{a}_i}_{ 2^\ekk }$, and a secret-shared label vector $\share{\vec{y}}_{ 2^\ekk }$.  \\
 \textbf{Output:}  A secret-shared sample-threshold vector $\share{\vv{\mathit{spth}}_i}_{ 2^\ekk }$ and a secret-shared modified Gini impurity vector $\share{\vv{\mathit{gini}}_i}_{ 2^\ekk }$.  \\
 \end{flushleft}

 \begin{algorithmic}[1]
    \STATE $ \share{\vv{\mathit{gini}}_i}_{ 2^\ekk } = \mathit{ModifiedGini}(\share{\vec{g}}_{ 2^\ekk }, \share{\vec{y}}_{ 2^\ekk })$.
    \STATE $ \share{\vec{\mathit{t}}[j]}_{ 2^\ekk } = (\share{\vec{a}_i[j]}_{ 2^\ekk } + \share{\vec{a}_i[j+1]}_{ 2^\ekk })\ /\ 2$ for each $j \in [0, n-1)$.
    \STATE $ \share{\vec{\mathit{t}}[n-1]}_{ 2^\ekk } = \share{\vec{a}_i[n-1]}_{ 2^\ekk }$.
    \STATE $ \share{\vec{p}[j]}_{ 2^\ekk } = \share{\vec{g}[j+1]}_{ 2^\ekk }$ \ or \  $(\share{\vec{a}_i[j]}_{ 2^\ekk } \overset{?}{=} \share{\vec{a}_i[j+1]}_{ 2^\ekk })$ for each $j \in [0, n-2)$.
    \STATE $ \share{\vec{p}[n-1]}_{ 2^\ekk } = \share{1}_{2^\ekk}$.
    \STATE $ \share{\vv{\mathit{gini}}_i}_{ 2^\ekk } = \mathit{MinValue} * \share{\vec{p}}_{ 2^\ekk } + \share{\vv{\mathit{gini}}_i}_{ 2^\ekk } * (1 - \share{\vec{p}}_{ 2^\ekk })$. 
  
    \STATE $ \share{\vv{\mathit{gini}}_i}_{ 2^\ekk }$,$ \share{\vv{\mathit{spth}}_i}_{ 2^\ekk } $ $=$ $ \mathit{GroupMax}(\share{\vec{g}}_{ 2^\ekk }$,$\share{\vv{\mathit{gini}}_i}_{ 2^\ekk }$,$\share{\vec{\mathit{t}}}_{ 2^\ekk })$.

 \end{algorithmic}
\end{algorithm}

\begin{algorithm}
    \LinesNumbered
    \small
    \caption{$\mathit{TestSamples}$}
    \label{pro:testsamples}
    \begin{flushleft}
 \textbf{Input:}  $m$ secret-shared attribute vectors, $\share{\vec{a}_0}_{ 2^\ekk },\cdots,\share{\vec{a}_{m-1}}_{ 2^\ekk }$,  a secret-shared sample-attribute  vector, $\share{\vv{\mathit{spat}}}_{ 2^\ekk }$,and  a secret-shared sample-threshold vector, $\share{\vv{\mathit{spth}}}_{ 2^\ekk }$.  \\
 \textbf{Output:} A secret-shared comparison result vector $\share{\vec{b}}_{ 2^\ekk }$.\\
 \end{flushleft}

 \begin{algorithmic}[1]
    \FOR{  each $ i \in [0, m-1]$ in parallel  }  
        \STATE $\share{\vv{eq}_i}_{ 2^\ekk } =( \share{\vv{\mathit{spat}}}_{ 2^\ekk } \overset{?}{=} i) $.
    \ENDFOR
    \STATE $\share{\vec{x}}_{ 2^\ekk } = \sum_{i=0}^{m-1}\share{\vec{a}_i}_{ 2^\ekk }*\share{\vv{eq}_i}_{ 2^\ekk }$.
    \STATE $\share{\vec{b}}_{ 2^\ekk } =  \share{\vec{x}}_{ 2^\ekk } < \share{\vv{\mathit{spth}}}_{ 2^\ekk } $.
\end{algorithmic}
\end{algorithm}

The protocol \textit{TestSamples} used to compute the comparison results between samples and the split threshold is shown in Protocol~\ref{pro:testsamples}. The parties in this protocol input $m$ secret-shared attribute vectors, $\share{\vec{a}_0}_{ 2^\ekk }$,$\cdots$,$\share{\vec{a}_{m-1}}_{ 2^\ekk }$,  a secret-shared sample-attribute  vector $\share{\vv{\mathit{spat}}}_{ 2^\ekk }$, and a secret-shared sample-threshold vector $\share{\vv{\mathit{spth}}}_{ 2^\ekk }$. They get a secret-shared comparison result vector $\share{\vec{b}}_{ 2^\ekk }$ as the output. Initially, the parties determine the attribute for testing the samples based on $\share{\vv{\mathit{spat}}}_{ 2^\ekk }$ and place the corresponding attribute values in a secret-shared vector $\share{\vec{x}}_{2^\ekk}$ (Line 1-4). Subsequently, the parties compute the secret-shared comparison results by comparing the secret-shared vector $\share{\vec{x}}_{2^\ekk}$ with the secret-shared sample-threshold vector $\share{\vv{\mathit{spth}}}_{ 2^\ekk }$ (Line 5). 

Note that this protocol is also the same as the protocol proposed by Hamada et al.~\cite{hamada2021efficient}.

The protocol \textit{TrainLeafLayer} used to train a leaf layer is shown in Protocol~\ref{pro:trainleaflayer}.  The parties in this protocol input an integer $k$ that denotes the height of the leaf layer, a secret-shared permutation $\share{\vec{\pi}}_{ 2^\ekk }$, a secret-shared sample-node vector $\share{\vv{\mathit{spnd}}}_{ 2^\ekk }$, and a secret-shared label vector $\share{\vec{y}}_{ 2^\ekk }$. The parties get a secret-shared leaf layer $\share{layer}_{ 2^\ekk }$ as the output. 

The protocol \textit{TrainLeafLayer}  consists of three stages: (1) The first stage (Line 1-4): the parties compute the secret-shared group flag vector $\share{\vec{g}}_{ 2^\ekk }$ using the same way in the protocol \textit{TrainInternalLayer} (Protocol~\ref{pro:traininternallayer}). (2) The second stage (Line 5-9): the parties compute the secret-shared sample-label vector $\share{\vv{splb}}_{ 2^\ekk }$. The equation $\vv{splb}[j]=i$ represents that the predicted label of the node to which the $j$-th sample belongs is $i$.  As the predicted label of a leaf node is the most common one of the samples belonging to it, the parties count the number of each label in each node by calling the protocol \textit{GroupSum} (introduced in Section~\ref{sec.group}) (Line 5-8). Then the parties determine which label is the most common one by calling the protocol \textit{VectMax} (introduced in Section~\ref{sec.vectmax}) (Line 9). (3) The third stage (Line 9): the parties call the protocol \textit{FormatLayer} (Protocol~\ref{pro:formatlayer}) to get the secret-shared layer. The protocol \textit{FormatLayer} is used to remove
the redundant values from the input vectors to leave only one nid
and one predicted label for each node.

\begin{algorithm}
    \LinesNumbered
    \small
    \caption{$\mathit{TrainLeafLayer}$}
    \label{pro:trainleaflayer}
    \begin{flushleft}
 \textbf{Input:}  An integer $k$ that represents the height of the leaf layer, a secret-shared permutation $\share{\vec{\pi}}_{ 2^\ekk }$,  a secret-shared sample-node vector $\share{\vv{\mathit{spnd}}}_{ 2^\ekk }$, and a secret-shared label vector  $\share{\vec{y}}_{ 2^\ekk }$.  \\
 \textbf{Output:} A secret-shared leaf layer $\share{layer}$.\\
 \end{flushleft}

 \begin{algorithmic}[1]
     \STATE $ \share{\vec{y'}}_{ 2^\ekk } = \mathit{ApplyPerm}(\share{\vec{\pi}}_{ 2^\ekk }, \share{\vec{y}}_{ 2^\ekk })$.
     \STATE $ \share{\vv{\mathit{spnd'}}}_{ 2^\ekk } = \mathit{ApplyPerm}(\share{\vec{\pi}}_{ 2^\ekk }, \share{\vv{\mathit{spnd}}}_{ 2^\ekk })$.
    \STATE $\share{\vec{g}[0]}_{ 2^\ekk } = \share{1}_{2^\ekk}$ 
    \STATE $\share{\vec{g}[j]}_{ 2^\ekk } = (\share{\vv{\mathit{spnd'}}[j-1]}_{ 2^\ekk } \neq \share{\vv{\mathit{spnd'}}[j]}_{ 2^\ekk })$ for each $j \in [1, n-1]$.

    \FOR{ each $ i \in [0, v-1]$ in parallel } 
            \STATE $\share{\vec{y_i}}_{ 2^\ekk } =  \share{\vec{y'}}_{ 2^\ekk } \overset{?}{=} i$.
            \STATE $\share{\vv{cnt}_i}_{ 2^\ekk } = \mathit{GroupSum}(\share{\vec{g}}_{ 2^\ekk }, \share{\vec{y_i}}_{ 2^\ekk })$.
    \ENDFOR
    \STATE $\share{\vv{splb}[j]}_{ 2^\ekk } = \mathit{VectMax}([\share{\vv{cnt}_0[j]}_{ 2^\ekk }, \cdots, \share{\vv{cnt}_{v-1}[j]}_{ 2^\ekk }], $\\$[0, \cdots, v-1])$ for each $j \in [0, n-1]$.

    \STATE $\share{layer}_{ 2^\ekk } = \mathit{FormatLayer}(k, \share{\vec{g}}_{ 2^\ekk }, \share{\vv{\mathit{spnd'}}}_{ 2^\ekk }, \share{\vv{splb}}_{ 2^\ekk })$.
\end{algorithmic}
\end{algorithm}

\begin{algorithm}
    \LinesNumbered
    \small
    \caption{$\mathit{FormatLayer}$}
    \label{pro:formatlayer}
    \begin{flushleft}
 \textbf{Input:}  An integer $k$ that represents the height of the layer required to be formatted,  a secret-shared group flag vector $\share{\vec{g}}_{ 2^\ekk }$, and $c$ secret-shared vectors $\share{\vec{w}_0}_{ 2^\ekk }, \cdots,\share{\vec{w}_{c-1}}_{ 2^\ekk }$ to be formatted.  \\
 \textbf{Output:} $c$ formatted secret-shared vectors $\share{\vec{u}_0}_{ 2^\ekk }, \cdots,\share{\vec{u}_{c-1}}_{ 2^\ekk }$.\\
 \end{flushleft}

 \begin{algorithmic}[1]
     \STATE $\share{\vec{\alpha}}_{ 2^\ekk } = \mathit{GenPermFromBit}(1 - \share{\vec{g}}_{ 2^\ekk })$.
    \FOR{ each $ i \in [0, c-1]$ in parallel } 
        \STATE $\share{\vec{u}_i}_{ 2^\ekk } = \mathit{ApplyPerm}(\share{\vec{\alpha}}_{ 2^\ekk }, \share{\vec{w}_i}_{ 2^\ekk })$.
       \STATE $\share{\vec{u}_i}_{ 2^\ekk } = \share{\vec{u}_i[0: \mathit{min}(2^k, n)]}_{ 2^\ekk }$. // retain the first $\mathit{min}(2^k, n)$ elements of $\share{\vec{u}_i}_{ 2^\ekk }$.

    \ENDFOR
\end{algorithmic}
\end{algorithm}
The protocol $\mathit{FormatLayer}$ used to format a layer is shown in Protocol~\ref{pro:formatlayer}. The parties in this protocol input an integer $k$ that denotes the height of the current layer required to be formatted, a secret-shared group flag vector $\share{\vec{g}}$, and $c$ secret-shared vectors $\share{\vec{w_0}}$, $\cdots$, $\share{\vec{w}_{c-1}}$. The parties get $c$ formatted secret-shared vectors $\share{\vec{u_0}}, \cdots, \share{\vec{u}_{c-1}}$ as the output. The protocol $\mathit{FormatLayer}$ aims to remove redundant values from the given vectors. The given vectors include the secret-shared sample-node vector, the secret-shared sample-attribute vector, the secret-shared sample-threshold vector, and the secret-shared label vector, whose elements are identical within each node. Thus it suffices to retain only one element per node. To achieve this, the parties move the first element of each group to the front of the vectors and retain only the $min(2^k, n)$ elements, as the $k$-th layer has at most $2^k$ nodes. As a result of this operation, the output decision tree is typically a complete binary tree, except when the number of training samples is fewer than the node number in some layers of a complete binary tree.









\section{Example of Training Steps}
\label{app.example}

\begin{figure}[htbp]
    \centering
    \includegraphics[width=0.46\textwidth]{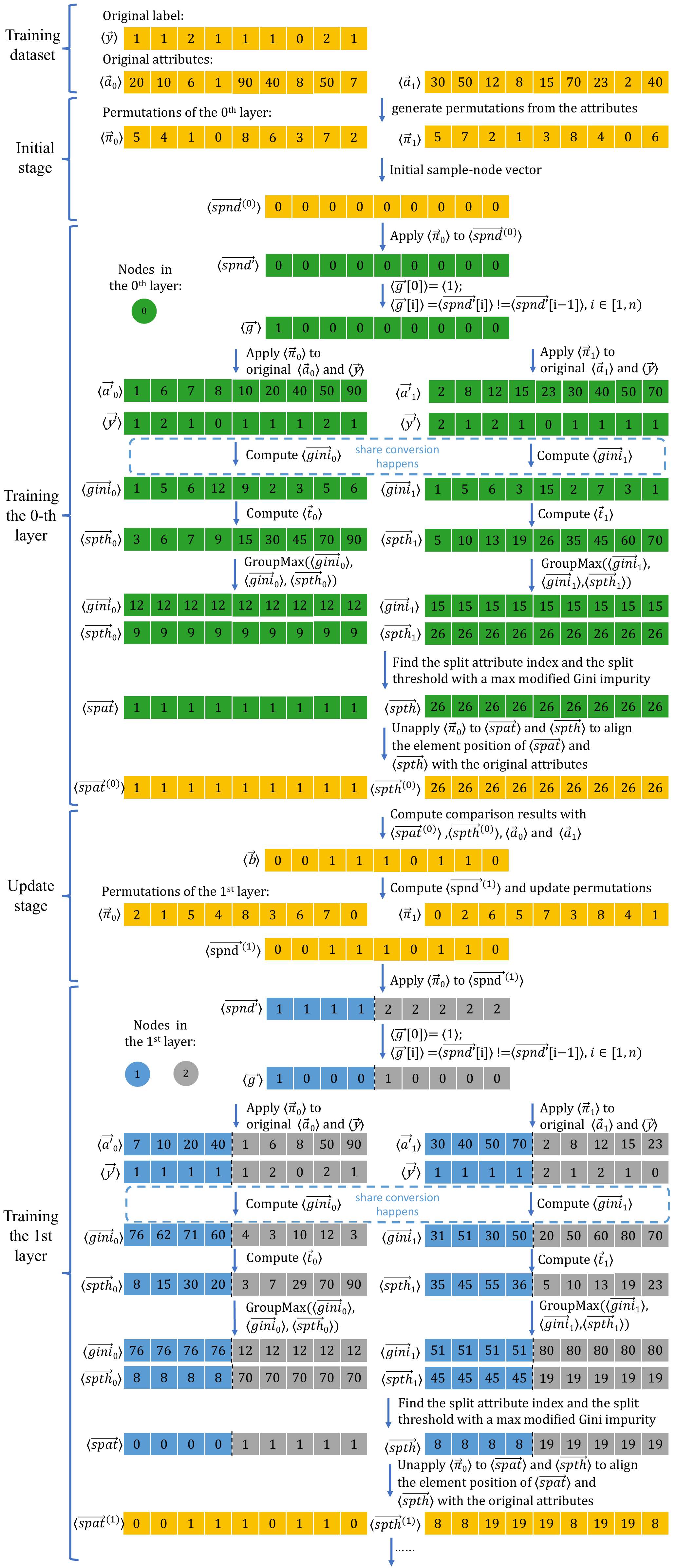}
    \caption{An example to show the key steps to train a decision tree with height more than two.  Vectors colored orange indicate that elements are stored in their original positions. Vectors colored green, blue, and gray indicate that elements belonging to the same nodes are stored consecutively.  Additionally, we use simulated values for the modified Gini vectors, because the real values contain too many digits. }

    \label{fig.trainingprocess}

\end{figure}

As is shown in Figure~\ref{fig.trainingprocess}, we illustrate the key steps of the training process. The example considers a training dataset containing nine samples, each with one label and two attributes.

 In the initial stage: the parties generate secret-shared permutations from the attribute vectors. These permutations could be directly used to sort the attributes and labels according to the corresponding attributes within each node of the 0-th layer. That is because this layer contains only a single node, with all samples belonging to it. Additionally, the parties initialize a secret-shared sample-node vector $\share{\vv{\mathit{spnd}}^{(0)}}$. 
 
 In the stage of training the 0-th layer: the parties first apply one of the secret-shared permutations (without loss of generality, $\share{\vec{\pi}_0}_{2^\ekk}$ is chosen) to $\share{\vv{\mathit{spnd}}^{(0)}}$, resulting in a new sample-node vector $\share{\vv{\mathit{spnd'}}}$. This step ensures that elements with the values in $\share{\vv{\mathit{spnd'}}}$ are stored consecutively. Note that this step could be omitted in the 0-th layer, since the 0-th layer only contains one node. The parties then use $\share{\vv{\mathit{spnd'}}}$ to compute the secret-shared group flag vector $\share{\vec{g}}$. Following this, they apply the secret-shared permutations to the secret-shared attribute vectors and the secret-shared label vector, obtaining new secret-shared attribute vectors and a new secret-shared label vector, whose elements are sorted according to the corresponding attribute vectors within each node. The parties proceed to compute the modified Gini impurity and the split thresholds of each split point for each attribute. Note that the communication optimization based on our proposed share conversion protocol is performed during the computation for the modified Gini impurity. Subsequently, the parties find a maximum modified Gini impurity and its corresponding threshold for each attribute by calling the protocol \textit{GroupMax} (introduced in Section~\ref{sec.group}). Then, the parties find the split attribute index and the split threshold with a maximum modified Gini impurity across the attributes. Finally, the parties align the element positions of $\share{\vv{\mathit{spat}}}$ and $\share{\vv{\mathit{spth}}}$ with those in the original attribute vectors by calling the protocol \textit{UnApply} (introduced in Section~\ref{sec.sort}). 
 
 In the update stage: the parties compute the comparison results between the samples and the split thresholds. Based on these results, the parties compute the secret-shared sample-node vector of the next layer and update the secret-shared permutations for training the next layer.

\section{\fname in Two-Party Scenarios}

\subsection{Share Conversion Protocol for Two-Party Scenarios}
\label{appendix.twopartyextend}

The protocol \textit{ConvertShareTwoParty} designed for share conversion in two-party scenarios is shown in Protocol~\ref{pro:twopartyconvert}. Here, we use the notation $\additiveshare{x}$ to denote the additive shares of a secret $x$, i.e. $P_0$ holds $\additiveshare{x}^0$ and $P_1$ holds $\additiveshare{x}^1$. The parties in this protocol input an additive share $\additiveshare{x}_{ 2^\ekk }$ on $\mathbb{Z}_{2^\ekk}$ and two public integers $\ekk$ and $\ell$ ($\ekk < \ell$), and they get a new additive share $\additiveshare{x'}_{2^\ell}$ on $\mathbb{Z}_{2^\ell}$ as the output. The implementation is similar to the three-party version, with only differences in computations of $d_0$, $d_1$, $b_0$, and $b_1$. Concretely, in this protocol, $P_0$ locally computes $d_0 =  \additiveshare{x}^0_{ 2^\ekk }$ and $b_0 = (\mathit{truncd}_0 \wedge 1) \oplus 
 \additiveshare{r}^0_2 $.  $P_1$ locally computes $d_1 =  \additiveshare{x}^1_{ 2^\ekk }$ and $b_1 = (\mathit{truncd}_1 \wedge 1) \oplus 
 \additiveshare{r}^1_2 $. while in the protocol \textit{ConvertShare}, $P_0$  locally computes $d_0 =  (\additiveshare{x}^0_{ 2^\ekk } + \additiveshare{x}^1_{ 2^\ekk }) \ mod \ 2^\ekk$ and $b_0 = (\mathit{truncd}_0 \wedge 1) \oplus \additiveshare{r}^0_2 \oplus \additiveshare{r}^1_2$.  $P_1$  locally computes $d_1 =  \additiveshare{x}^2_{ 2^\ekk }$ and $b_1 = (\mathit{truncd}_1 \wedge 1) \oplus \additiveshare{r}^2_2$.

We analyze the security of this protocol in Appendix~\ref{app.securityanalysis}.

\begin{algorithm}
    \LinesNumbered
    \small
    \caption{$\mathit{ConvertShareTwoParty}$}
    \label{pro:twopartyconvert}
    \begin{flushleft}
  \textbf{Input:}    $\additiveshare{x}_{ 2^\ekk } (x \in [0, 2^{\ekk-1}) )$, i.e.  $P_i$ inputs $\additiveshare{x}^i_{ 2^\ekk }$, and two public integers $\ekk$ and $\ell$ ($\ekk < \ell$).  \\
 \textbf{Output:}  $\additiveshare{x'}_{2^\ell}$, i.e.  $P_i$ gets  $[x']^i_{2^\ell}$ , with $x' = x$.  \\
 \textbf{Offline:}  $P_0, P_1$ generate a daBit together, so each $P_i$ holds $\additiveshare{r}^i_2$, and $\additiveshare{r}^i_{2^\ell}$  ( $r = 0 \ or \ 1$ ). \\
 \end{flushleft}

 \begin{algorithmic}[1]
 
     \STATE $P_0$ locally computes $d_0 =  \additiveshare{x}^0_{ 2^\ekk }$.  $P_1$ locally computes $d_1 = \additiveshare{x}^1_{ 2^\ekk }$.
     \STATE $P_0$ locally computes $\mathit{truncd}_0 = d_0 \gg (\ekk-1)$.  $P_1$ locally computes  $ \mathit{truncd}_1 = -((-d_1) \gg (\ekk-1)) $.
    \STATE $P_0$ shares  $d_0$ and $\mathit{truncd}_0$ on $\mathbb{Z}_{2^\ell}$. $P_1$ shares $d_1$ and $\mathit{truncd}_1$ on $\mathbb{Z}_{2^\ell}$. 
    \STATE $\additiveshare{d}_{2^\ell} = \additiveshare{d_0}_{2^\ell} + \additiveshare{d_1}_{2^\ell}  $.
    \STATE $\additiveshare{\mathit{truncsum}}_{2^\ell} = \additiveshare{\mathit{truncd}_0}_{2^\ell} + \additiveshare{\mathit{truncd}_1}_{2^\ell}  $.
    
     \STATE  $P_0$ locally computes $b_0 = (\mathit{truncd}_0 \wedge 1) \oplus 
 \additiveshare{r}^0_2 $.  $P_1$ locally computes $ b_1 = (\mathit{truncd}_1 \wedge 1) \oplus \additiveshare{r}^1_2$.
    \STATE $P_0$  sends $b_0$ to $P_1$. $P_1$  and sends $b_1$ to $P_0$. 
    \STATE The parties both locally compute $b = b_0 \oplus b_1  $.
    \STATE $\additiveshare{bit}_{2^\ell} = b + \additiveshare{r}_{2^\ell} - 2 * b * \additiveshare{r}_{2^\ell}$.

    \STATE $\additiveshare{\mathit{ovfl}}_{2^\ell} = (\additiveshare{\mathit{truncsum}}_{2^\ell} - \additiveshare{bit}_{2^\ell}) * 2^{\ekk-1}  $.

    \STATE $\additiveshare{x'}_{2^\ell} = \additiveshare{d}_{2^\ell} - \additiveshare{\mathit{ovfl}}_{2^\ell} $.
 \end{algorithmic}

\end{algorithm}

\subsection{Experimental Results of Two-Party \fname}
\label{appendix.twopartyexperiment}
We compare the efficiency of two-party \fname with two state-of-the-art two-party frameworks~\cite{liu2020towards, akavia2022privacy}. (1) Liu et al.'s framework~\cite{liu2020towards} is based on secret sharing and homomorphic encryption. It offers three privacy levels. Higher privacy levels provide stronger privacy guarantees. We compare two-party \fname with the framework~\cite{liu2020towards} at its highest privacy level, as two-party \fname provides stronger privacy guarantees. Specifically, even at the highest privacy level, the framework~\cite{liu2020towards} still leaks Gini impurity and the decision tree structure, while two-party \fname does not leak any private information during the training process. Since this framework is not open-sourced, we rely on the experimental results, which are performed on two laptops with 4-core 2.6GHz Intel Core i7-6700HQ CPUs and 8GB of RAM, reported in their paper~\cite{liu2020towards}. (2) Akavia et al.'s framework~\cite{akavia2022privacy} is based on the CKKS homomorphic encryption scheme~\cite{cheon2017homomorphic} implemented in Microsoft SEAL v3.3.2~\cite{sealcrypto}. The framework involves a client and a server, with the client outsourcing encrypted data to the server to train decision trees.  As this framework is also not open-sourced, we use the experimental results, which were conducted on an AWS x1.16xlarge server equipped with 32-core server 2.3GHZ and 976GB of RAM, from their paper~\cite{akavia2022privacy}. 

In terms of hardware settings, our setup is worse than that used in Akavia et al.'s  framework~\cite{akavia2022privacy}. When compared to the framework by Liu et al.\cite{liu2020towards}, our core usage is the same as theirs, with each party utilizing up to 4 cores. Although our machines have more memory, the actual memory usage for evaluating the three datasets (Kohkiloyeh, Diagnosis, and Tic-tac-toe) in two-party \fname does not exceed 200 MB of RAM.

\noindent\textbf{Results.} We show the experimental results in Table~\ref{tab.twopartyefficiency}. Note that the literatures~\cite{liu2020towards, akavia2022privacy} only report the experimental results obtained in the LAN setting (their concrete RTT and bandwidth are unknown). We consider the experimental results of the framework~\cite{akavia2022privacy} in the LAN setting to be representative of the WAN setting as well, because the framework~\cite{akavia2022privacy} is totally based on homomorphic encryption and its performance is usually unaffected by the network setting. The experimental results show that two-party \fname significantly outperforms the framework~\cite{liu2020towards} by $14.3\times \sim 36.9\times$ in the LAN setting, and outperforms the framework~\cite{akavia2022privacy} by $338.2\times \sim 1,362.3\times$ in the LAN setting and by $4.7\times \sim 6.4\times$ in the WAN setting. The improvement is primarily due to two facts: (1) the computations of two-party \fname are totally performed based on additive secret sharing, which is much more computation efficient compared to homomorphic encryption. Such a technical route makes two-party \fname more efficient than the two frameworks~\cite{liu2020towards, akavia2022privacy}, especially in the LAN setting. (2) with the two communication optimizations, the communication overhead of \fname should be low, ensuring its efficiency even in the WAN setting.

\begin{table}[ht]
\center
\caption{Online training time (seconds) of two-party \fname vs. the two-party frameworks~\cite{liu2020towards, akavia2022privacy} for training a decision tree. Since the tree height for the framework~\cite{liu2020towards} is not specified, we trained a decision tree of height six for \fname on the Kohkiloyeh, Diagnosis, and Tic-tac-toe datasets. For the datasets Iris, Wine, and Cancer, we matched the tree height and sample number with those used in the literature~\cite{akavia2022privacy}. Concretely, the tree height for the three datasets is set to four, and the sample number is set to 100, 119, and 381, respectively. The notation `-' indicates the absence of experimental results. }

\scalebox{0.9}{
\begin{tabular}{|c|ccc|cc|}
\hline
\multirow{2}{*}{Dataset} & \multicolumn{3}{c|}{LAN}                                                                                                                                                 & \multicolumn{2}{c|}{WAN}                                                                                                    \\ \cline{2-6} 
                         & \multicolumn{1}{c|}{\fname}                                                                      & \multicolumn{1}{c|}{\cite{liu2020towards}} & \cite{akavia2022privacy} & \multicolumn{1}{c|}{\fname}                                                                      & \cite{akavia2022privacy} \\ \hline
Kohkiloyeh               & \multicolumn{1}{c|}{\textbf{\begin{tabular}[c]{@{}c@{}}4.15\\ ($36.99 \times$)\end{tabular}}}    & \multicolumn{1}{c|}{153.53}                & -                        & \multicolumn{1}{c|}{\textbf{943.70}}                                                             & -                        \\ \hline
Diagnosis                & \multicolumn{1}{c|}{\textbf{\begin{tabular}[c]{@{}c@{}}4.26\\ ($14.31 \times$)\end{tabular}}}    & \multicolumn{1}{c|}{61}                    & -                        & \multicolumn{1}{c|}{\textbf{943.91}}                                                             & -                        \\ \hline
Tic-tac-toe              & \multicolumn{1}{c|}{\textbf{\begin{tabular}[c]{@{}c@{}}56.52\\ ($23.67 \times$)\end{tabular}}}   & \multicolumn{1}{c|}{1,338}                 & -                        & \multicolumn{1}{c|}{\textbf{2,056.0}}                                                            & -                        \\ \hline
Iris                     & \multicolumn{1}{c|}{\textbf{\begin{tabular}[c]{@{}c@{}}2.07\\ ($1,362.31 \times$)\end{tabular}}} & \multicolumn{1}{c|}{-}                     & 2,820                    & \multicolumn{1}{c|}{\textbf{\begin{tabular}[c]{@{}c@{}}435.40\\ ($6.47 \times$)\end{tabular}}}   & 2,820                    \\ \hline
Wine                     & \multicolumn{1}{c|}{\textbf{\begin{tabular}[c]{@{}c@{}}7.87\\ ($1,128.33 \times$)\end{tabular}}} & \multicolumn{1}{c|}{-}                     & 8,880                    & \multicolumn{1}{c|}{\textbf{\begin{tabular}[c]{@{}c@{}}1,674.91\\ ($5.30 \times$)\end{tabular}}} & 8,880                    \\ \hline
Cancer                   & \multicolumn{1}{c|}{\textbf{\begin{tabular}[c]{@{}c@{}}49.31\\ ($338.26 \times$)\end{tabular}}}  & \multicolumn{1}{c|}{-}                     & 16,680                   & \multicolumn{1}{c|}{\textbf{\begin{tabular}[c]{@{}c@{}}3,533.82\\ ($4.72 \times$)\end{tabular}}} & 16,680                   \\ \hline
\end{tabular}
}
\label{tab.twopartyefficiency}

\end{table}

\section{Security Analysis For Conversion Protocols}

\label{app.securityanalysis}

We analyze the security of the protocols \textit{ConvertShare} (Protocol~\ref{pro:convertshare})  and \textit{TwoPartyConvertShare} (Protocol~\ref{pro:twopartyconvert}) using the standard real/ideal world paradigm.

\begin{proof}
Let the semi-honest adversary $\mathcal{A}$ corrupt at most one party, we now present the steps of the idea-world adversary (simulator) $\mathcal{S}$ for $\mathcal{A}$. Our simulator $\mathcal{S}$ for  \textit{ConvertShare} (Protocol~\ref{pro:convertshare}),  and \textit{ConvertShareTwoParty} (Protocol~\ref{pro:twopartyconvert}) is constructed as follows:

\noindent\textbf{Security for \textit{ConvertShare} (Protocol~\ref{pro:convertshare})}:
For the offline phase, we use the generation process of the daBit~\cite{damgaard2019new} in a black-box manner. Therefore, the simulation follows the same in the study~\cite{damgaard2019new}. For line 3 and 7 in Protocol~\ref{pro:convertshare}, which are the only two steps that require interaction between parties, we analyze it case by case:
(1) If $\mathcal{A}$ corrupts $P_0$, $\mathcal{S}$ receives $\additiveshare{d_0}^1_{2^\ell}$, $\additiveshare{d_0}^2_{2^\ell}$, $\additiveshare{truncd_0}^1_{2^\ell}$, $\additiveshare{truncd_0}^2_{2^\ell}$ and $b_0$, from $\mathcal{A}$ on behalf of $P_1$, and receives $\additiveshare{d_0}^2_{2^\ell}$, $\additiveshare{d_0}^0_{2^\ell}$, $\additiveshare{truncd_0}^2_{2^\ell}$, $\additiveshare{truncd_0}^0_{2^\ell}$ and $b_0$, from $\mathcal{A}$ on behalf of $P_2$. Then $\mathcal{S}$ selects random values to simulate  $\additiveshare{d_1}^0_{2^\ell}$, $\additiveshare{d_1}^1_{2^\ell}$, $\additiveshare{truncd_1}^0_{2^\ell}$, $\additiveshare{truncd_1}^1_{2^\ell}$ and $b_1$, and sends them to $\mathcal{A}$ on behalf of $P_1$.
(2) If $\mathcal{A}$ corrupts $P_1$, $\mathcal{S}$ receives $\additiveshare{d_1}^0_{2^\ell}$, $\additiveshare{d_1}^1_{2^\ell}$, $\additiveshare{truncd_1}^0_{2^\ell}$, $\additiveshare{truncd_1}^1_{2^\ell}$ and $b_1$, from $\mathcal{A}$ on behalf of $P_0$, and receives $\additiveshare{d_1}^2_{2^\ell}$, $\additiveshare{d_1}^0_{2^\ell}$, $\additiveshare{truncd_1}^2_{2^\ell}$, $\additiveshare{truncd_1}^0_{2^\ell}$ and $b_1$, from $\mathcal{A}$ on behalf of $P_2$. Then $\mathcal{S}$ selects random values to simulate  $\additiveshare{d_1}^0_{2^\ell}$, $\additiveshare{d_1}^1_{2^\ell}$, $\additiveshare{truncd_1}^0_{2^\ell}$, $\additiveshare{truncd_1}^1_{2^\ell}$ and $b_1$, and sends them to $\mathcal{A}$ on behalf of $P_0$.
(3) If $\mathcal{A}$ corrupts $P_2$,  $\mathcal{S}$ selects random values to simulate  $\additiveshare{d_1}^0_{2^\ell}$, $\additiveshare{d_1}^1_{2^\ell}$, $\additiveshare{truncd_1}^0_{2^\ell}$, $\additiveshare{truncd_1}^1_{2^\ell}$ and $b_1$, and sends them to $\mathcal{A}$ on behalf of $P_0$. Besides, $\mathcal{S}$ selects random values to simulate  $\additiveshare{d_1}^0_{2^\ell}$, $\additiveshare{d_1}^1_{2^\ell}$, $\additiveshare{truncd_1}^0_{2^\ell}$, $\additiveshare{truncd_1}^1_{2^\ell}$ and $b_1$, and sends them to $\mathcal{A}$ on behalf of $P_1$.

\noindent\textbf{Security for \textit{ConvertShareTwoParty} (Protocol~\ref{pro:twopartyconvert})}:
For the offline phase, we use the generation process of the daBit~\cite{damgaard2019new} in a black-box manner. Therefore, the simulation follows the same in the study~\cite{damgaard2019new}. For line 3 and 7 in Protocol~\ref{pro:convertshare}, which are the only two steps that require interaction between parties, we analyze it case by case:
(1) If $\mathcal{A}$ corrupts $P_0$, $\mathcal{S}$ receives $\additiveshare{d_0}^1_{2^\ell}$,  $\additiveshare{truncd_0}^1_{2^\ell}$ and $b_0$, from $\mathcal{A}$ on behalf of $P_1$. Then $\mathcal{S}$ selects random values to simulate  $\additiveshare{d_1}^0_{2^\ell}$, $\additiveshare{truncd_1}^0_{2^\ell}$ and $b_1$, and sends them to $\mathcal{A}$ on behalf of $P_1$.
(2) If $\mathcal{A}$ corrupts $P_1$, $\mathcal{S}$ receives $\additiveshare{d_1}^0_{2^\ell}$, $\additiveshare{truncd_1}^0_{2^\ell}$ and $b_1$, from $\mathcal{A}$ on behalf of $P_0$. Then $\mathcal{S}$ selects random values to simulate  $\additiveshare{d_1}^0_{2^\ell}$, $\additiveshare{truncd_1}^0_{2^\ell}$ and $b_1$, and sends them to $\mathcal{A}$ on behalf of $P_0$.

Since all the messages sent and received in the protocol are uniformly random values in both the real protocol and the simulation,  the $\mathcal{A}$'views in the real and ideal worlds are both identically distributed and indistinguishable. This concludes the proof.
\end{proof}

\section{Details of communication complexity Analysis}

\label{app.communication_complexity}

We first present the communication complexity of basic primitives introduced in Section~\ref{sec.mpc} and Appendix~\ref{app.basic_operations}.

\begin{itemize}[leftmargin=*]
    \item \textit{Basic Operations.} Assuming the basic operations are performed on a ring $\mathbb{Z}_{2^\ekk}$, their communication complexity is as follows: (1) Secure addition with constant, secure multiplication with constant, and secure addition usually can be performed without communication. (2) Secure multiplication and secure probabilistic truncation both usually require a communication size of $O(\ekk)$ in $O(1)$ communication rounds. (3) Secure comparison and equality test usually require a communication size of $O(\ekk \log \ekk)$ in $O(\ekk)$ communication rounds. (4) Secure division usually requires a communication size of $O(\ekk \log f)$ in $O(\log f)$ communication rounds, where $f$ is the bit length of the decimal part and set to be $2 \lceil \log n \rceil$ in our evaluation.

    \item \textit{Vector Max Protocol Based on RSS.} Assuming the protocol \textit{VectMax}~\cite{hamada2021efficient} is performed on a ring $\mathbb{Z}_{2^\ekk}$ and its input secret-shared vectors are of size $n$, it requires a communication size of $O(n\ekk)$ bits in $O(\log n)$ communication rounds.

    \item \textit{Secure Radix Sort Protocols Based on RSS.} Assuming the secure radix sort protocols~\cite{chida2019efficient} Based on RSS are all performed on a ring $\mathbb{Z}_{2^\ekk}$ and their input secret-shared vectors are of size $n$, the communication complexity of the protocols is as follows: the protocols \textit{GenPerm} and \textit{BitVecDec} both require a communication size of $O(n\ekk^2)$ bits in $O(\ekk)$ communication rounds.  The protocols \textit{GenPermByBit}, \textit{ApplyPerm}, \textit{ComposePerms} and \textit{UnApplyPerm} all require a communication size of $O(n\ekk)$ bits in $O(1)$ communication rounds.

    \item \textit{Group-Wise Protocols Based on RSS.} Assuming group-wise protocols~\cite{hamada2021efficient} Based on RSS are all performed on a ring $\mathbb{Z}_{2^\ekk}$ and their input secret-shared vectors are of size $n$,  the communication complexity of the protocols is as follows: the protocols \textit{GroupSum} and \textit{GroupPrefixSum} both require a communication size of $O(n\ekk)$ bits in $O(1)$ communication rounds.    The protocol \textit{GroupMax} and its extension version require a communication size of  $O(n\log n\ekk \log \ekk)$ bits in $O(\log n \log \ekk)$ communication rounds \footnote{We show the communication complexity of optimized version of the group-wise protocols implemented in MP-SPDZ~\cite{keller2020mp}, which is used to implement \fname. Therefore, the communication complexity is inconsistent with the complexity in the literature~\cite{hamada2021efficient}.}. 

\end{itemize}

Next, we analyze the online communication complexity for each training protocol proposed by us as follows:

\begin{itemize}[leftmargin=*]

    \item \textit{TrainDecisionTree} (Protocol~\ref{pro:traindecisiontree}): In this protocol, the parties first call the protocol \textit{GenPerm} (protocol~\ref{pro:genperm}) $m$ times in parallel, which requires a communication size of $O(mn \ekk^2)$ bits in $O(\ekk)$ communication rounds. Next, the parties call the protocol \textit{TrainInternalLayer} (protocol~\ref{pro:traininternallayer}) $h$ times sequentially,  which requires a communication size of $O(hmn \log n \ekk \log \ekk )$ bits in $O(h\log n \log \ekk )$ communication rounds. Finally, the parties call the protocol \textit{TrainLeafLayer} (protocol~\ref{pro:trainleaflayer}), which requires a communication size of $O(vn\ekk)$ bits in $O(\log v)$ communication rounds. Therefore, the protocol \textit{TrainDecisionTree} requires a communication size of $O(hmn \log n \ekk \log \ekk + mn\ekk^2)$ bits in $O(h\log n \log \ekk + \ekk)$ communication rounds

    \item \textit{UpdatePerms} (Protocol~\ref{pro:updateperms}): In this protocol, the parties mainly call the protocols \textit{ApplyPerm} (introduced in Section~\ref{sec.sort}), \textit{GenPermByBit} (introduced in Section~\ref{sec.sort}), and \textit{ComposePerms} (introduced in Section~\ref{sec.sort}) $m$ times in parallel. Therefore, the protocol \textit{UpdatePerms} requires a communication size of $O(mn\ekk)$ bits in $O(1)$ communication rounds.

    \item \textit{TrainInternalLayer} (Protocol~\ref{pro:traininternallayer}): In this protocol, the parties mainly call the protocol \textit{AttributeWiseSplitSelection} (protocol~\ref{pro:attributewisesplitselection}) $m$ times in parallel,  which requires a communication size of $O(mn$ $\log n$ $\ekk \log \ekk)$ bits in $O(\log n \log \ekk)$ communication rounds, and call the protocol \textit{VectMax} (introduced in section~\ref{sec.vectmax}) $n$ times in parallel, which requires a communication size of $O(nm\ekk)$ bits in $O(\log m)$ communication rounds. Since $m$ is typically smaller than $n$, the protocol \textit{TrainInternalLayer} requires a communication size of $O(mn \log n \ekk \log \ekk)$ bits in $O(\log n \log \ekk)$ communication rounds.

    \item \textit{ComputeModifiedGini} (Protocol~\ref{pro:computemodifiedgini}): In this protocol, the parties mainly perform multiplication $v$ times in parallel, which requires a communication size of $O(nv\ell)$ bits in $O(1)$ communication rounds, and perform secure division operation one time, which requires a communication size of $O(n\log n \ell  )$ bits in $O(\log n)$ communication rounds. Therefore, the protocol \textit{ComputeModifiedGini} requires a communication size of $O(nv\ell + n\ell \log \ell)$ bits in $O(\log n)$ communication rounds.

        \item \textit{ConvertShare} (Protocol~\ref{pro:convertshare}): In this protocol, $P_0$ and $P_1$ collectively share four numbers on $\mathbb{Z}_{2^\ell}$ and transmit four bits. Note that, the sharing process and transmitting process can be performed in the same communication round. Besides, using the sharing method proposed in the study~\cite{mohassel2018aby3}, sharing a single number on $\mathbb{Z}_{2^\ell}$ only requires transmitting $\ell$ bits. Therefore, \textit{ConvertShare} requires a communication size of $4\ell + 4$ bits in one online communication round.

    \item \textit{AttributeWiseSplitSelection} (Protocol~\ref{pro:attributewisesplitselection}): In this protocol, the parties mainly call the protocol \textit{ComputeModifiedGini} (Protocol~\ref{pro:computemodifiedgini}), which requires a communication size of $O(nv\ell + n\ell \log \ell)$ bits in $O(\log n)$ communication rounds, and call the protocol \textit{GroupMax} (introduced in section~\ref{sec.group}), which requires a communication size of $O(n \log n \ekk \log \ekk)$ bits in $O(\log n \log \ekk)$ communication rounds. Therefore, the protocol \textit{AttributeWiseSplitSelection} requires a communication size of $O(n \log n \ekk \log \ekk)$ bits in $O(\log n \log \ekk)$ communication rounds.
    
    \item \textit{TestSamples} (Protocol~\ref{pro:testsamples}): In this protocol, the parties operate on $m$ secret-shared vectors in parallel. Therefore, the protocol \textit{TestSamples} requires a communication size of $O(mn\ekk)$ bits in $O(1)$ communication rounds.

    \item \textit{TrainLeafLayer} (Protocol~\ref{pro:trainleaflayer}): In this protocol, the parties mainly call the protocol \textit{ApplyPerm} (introduced in Section~\ref{sec.sort}) two times, which requires a communication size of $O(n\ekk)$ bits in $O(1)$ communication rounds, call the protocol \textit{GroupSum} (introduced in Section~\ref{sec.group}) $v$ times in parallel, which requires a communication size of $O(vn\ekk)$ bits in $O(1)$ communication rounds, call the protocol \textit{VectMax} (introduced in Section~\ref{sec.vectmax}) once, which requires a communication size of $O(vn\ekk)$ bits in $O(\log v)$ communication rounds, and call the protocol \textit{FormatLayer} (Protocol~\ref{pro:formatlayer}) one time, which requires a communication size of $O(n\ekk)$ bits in $O(1)$ communication rounds. Therefore, the protocol \textit{TrainLeafLayer} requires a communication size of $O(vn\ekk)$ bits in $O(\log v)$ communication rounds.
    
    \item \textit{FormatLayer} (Protocol~\ref{pro:formatlayer}): In this protocol, the parties mainly call the protocol \textit{GenPermFromBit} (introduced in Section~\ref{sec.sort}) once, which requires a communication size of $O(n\ekk)$ bits in $O(1)$ communication rounds, and call the protocol \textit{ApplyPerm} (introduced in Section~\ref{sec.sort}) $c$ times, which requires a communication size of $O(cn\ekk)$ bits in $O(1)$ communication rounds. Since $c$ is a small constant representing the number of vectors to be formatted, we disregard the factor $c$. Therefore,  the protocol \textit{FormatLayer} requires a communication size of $O(n\ekk)$ bits in $O(1)$ communication rounds.
    
\end{itemize}

\section{Proof of Theorem 1}

\label{app.proof-of-corollary}

We prove Theorem~\ref{theorem1} as follows.

\begin{proof}

We define 
$$d = d''* 2^c + d', d'' \in [0, 2^{\ekk+1-c}), d' \in [0, 2^c)$$
$$d_0 = d_0''* 2^c + d_0', d_0'' \in [0, 2^{\ekk-c}), d_0' \in [0, 2^c)$$

Since  $d_0 + d_1 = d$ , then $d_1 = d - d_0$. Since $d_0 \in [0, 2^\ekk]$ and $c < \ekk < \ell - 1 $, then $d \gg c = d''$
.

We prove the theorem case by case.

(1) If $d_0' \geq d'$: $d_0 - d = (d_0'' - d'') * 2^c + (d_0' - d')$. Then $(d_0 - d) \gg c = ((d_0'' - d'') * 2^c + (d_0' - d')) \gg c = (((d_0'' - d'') * 2^c) \gg c) + ((d_0' - d') \gg c) = d_0'' - d''$. Thus, $(d_0 \gg 
 c) + (-((-d_1)\gg c)) = d_0'' + (-((d_0 - d) \gg c)) = d_0'' + (- (d_0'' - d'')) = d''$

(2) If $d_0' < d'$: $d_0 - d = (d_0'' - d'' - 1) * 2^c + (d_0' + 2^c - d')$. Then $(d_0 - d) \gg c = ((d_0'' - d'' - 1) * 2^c + (d_0' + 2^c - d')) \gg c = (((d_0'' - d'' - 1) * 2^c) \gg c) + ((d_0' + 2^c - d') \gg c) = d_0'' - d'' - 1$. Thus, $(d_0 \gg c) + (-((-d_1)\gg c)) = d_0'' + (-((d_0 - d) \gg c)) = d_0'' + (-((d_0'' - d'' - 1) \gg c)) = d_0'' + (- (d_0'' - d'' - 1)) = d'' + 1$

In summary, $(d_0 \gg c) + (-((-d_1)\gg c)) = \lfloor {d / 2^c}  \rfloor + bit$, where $bit = 0 \ or \ 1$

\end{proof}

\end{document}